\documentclass[twocolumn,floatfix,amssymb,aps,superscriptaddress,letterpager]{revtex4-2}
\usepackage[colorlinks=true,urlcolor=blue,citecolor=blue,linkcolor=blue]{hyperref}
\usepackage{graphicx}
\usepackage{color}
\usepackage{epsfig}
\usepackage{amsmath}
\usepackage{amssymb}
\usepackage{mathrsfs}
\usepackage{bm}
\usepackage{subfigure}
\usepackage{float}
\usepackage{url}
\usepackage{enumerate}
\usepackage{changes}
\usepackage{soul}
\usepackage{braket}
\usepackage{comment}
\usepackage{multirow}
\usepackage{appendix}
\definecolor{violet}{rgb}{0.56,0.0,1.0}
\def\tbib#1{{\bf{#1}}}

\def\Tr{{\rm{Tr}}}
\def\ket#1{|#1\rangle}
\def\lr#1{\langle#1\rangle}
\def\tred#1{{\color{red}{#1}}}

\begin{document}
	
\hsize\textwidth\columnwidth\hsize\csname@twocolumnfalse\endcsname

\title{Efficient calculation of three-dimensional tensor networks}

\author{Li-Ping Yang}
\affiliation{Department of Physics, Chongqing University, Chongqing 401331, China}

\author{Y. F. Fu}
\affiliation{Department of Physics, Renmin University of China, Beijing 100872, China}

\author{Z. Y. Xie}
\email[]{qingtaoxie@ruc.edu.cn}
\affiliation{Department of Physics, Renmin University of China, Beijing 100872, China}

\author{T. Xiang}
\email[]{txiang@iphy.ac.cn}
\affiliation{Institute of Physics, Chinese Academy of Sciences, Beijing, 100190, China}
\affiliation{Beijing Academy of Quantum Information Sciences, Beijing 100193, China}
\affiliation{University of Chinese Academy of Sciences, Department of Physics, Beijing 100190, China}

\begin{abstract}
We have proposed an efficient algorithm to calculate physical quantities in the translational invariant three-dimensional tensor networks, which is particularly relevant to the study of the three-dimensional classical statistical models and the (2+1)-dimensional quantum lattice models. In the context of a classical model, we determine the partition function by solving the dominant eigenvalue problem of the transfer matrix, whose left and right dominant eigenvectors are represented by two projected entangled simplex states. These two projected entangled simplex states are not Hermitian conjugate to each other but are appropriately arranged so that their inner product can be computed much more efficiently than in the usual prescription. For the three-dimensional Ising model, the calculated internal energy and spontaneous magnetization agree with the published results in the literature. The possible improvement and extension to other models are also discussed.
\end{abstract}

\maketitle
\section{Introduction}
The many-body problem is one of the central problems in physics, and developing accurate and efficient numerical methods that can  
effectively handle the exponential growth of the corresponding Hilbert space has always been a great challenge, especially for quantum systems. Based on the idea of renormalization group and the tensor-network representation of partition functions and wave functions, tensor-network methods have evolved progressively to be an important member of many-body computational methods in recent years \cite{PEPS2004, SimBook2018, OrusReview}. In fact, due to the absence of sign problem and the ability to deal with two-dimensional systems, tensor networks are drawing increasing attention, and have been applied successfully to strongly-correlated electron systems \cite{tJ2014, Simons-XXS}, frustrated spin systems \cite{Corboz-PRL2014, WL-PRB2016, Liao-PRL2017, LQ-PRB2022, Ferrari-PRB2022}, statistical models \cite{HOTRG-PRB2012, CW-PRB2014, ZP-PRL2021}, topological order \cite{Eisert-PRB2017, ZGM-PRL2020, Frank-arXiv2017, RW-PBL2022}, quantum field theory \cite{CMPS2010, LQCD2013, CTNS2019}, machine learning \cite{ML2018, ML2020}, and quantum circuit simulation \cite{Yannick-RMP2022}, etc. Among the various tensor-network methods, imaginary time evolution is a highly efficient method to determine the ground state of low-dimensional quantum systems \cite{OrusReview, GV-TEBD, SU1D2007}.    

A central task in the application of tensor network states to 2+1 dimensional quantum lattice models is to contract a two-dimensional tensor network, which has a double-layer structure and bond dimension $D^2$, with $D$ the maximal virtual bond dimension of the tensor network representation of a quantum state \cite{OrusReview}. This is extremely costly, and thus although $D$ is expected to be larger to produce a more accurate result, it is limited to about $13$ in practical calculations \cite{PESS2014, LQ-PRB2022, RW-PBL2022}. There have been some efforts in recent years to solve this problem, such as employing symmetries \cite{HHZ-PRB2010, Orus-SU2}, combining Monte Carlo sampling \cite{RGMC-WHZ}, and the nested tensor network method \cite{NTN-PRB2017}. However, apparently the problem is not completely solved, and developing efficient algorithms to resolve this issue is still an important topic for tensor-network communities. 

Meanwhile, it is known that, in the formalism of path integral, the equilibrium quantum many-body problem in $d$ dimensions is similar to the classical many-body problem in $d$+1 dimensions \cite{SachdevBook-QPT2011, Suzuki-PTP1976, Nishino-JPSJ1995, Xiang-JPCM1996, Nig-ZPB1997, HOTRG-PRB2012}. Thus inspired by the success in one- and two-dimensional quantum systems, there are also some efforts to apply tensor-network methods to three-dimensional (3D) classical models. Actually, both the coarse-graining tensor renormalization group algorithms \cite{HOTRG-PRB2012, PYT-PA2017, Daisuke-arXiv2019, Daiki-PRB2020} and the transfer-matrix-based tensor-network state methods \cite{Nishino-JPSJ1998, Nishino-KWA, Nishino-TPVA, SGC-PLA2006, Orus-CTTRG2012, FV-PRE2018, LV-PRL2022} have been applied to these models or related previously. Although the transfer-matrix-based methods can work excellently in two-dimensional networks, they seem not so efficient in three-dimensions \cite{Nishino-JPSJ2022}. In the literature, in order to obtain reliable results, the variational optimization procedure is indispensable, thus due to the high computational cost, as will be discussed later, the bond dimension of the involved tensor-network states and the accuracy are rather limited, and some auxiliary techniques are usually employed further to analyze the data. Especially, to our best knowledge, in practice, whether a simple yet efficient algorithm analogous to the simple-update algorithm \cite{SU2D2008} in two-dimensional quantum lattice models can be developed for 3D classical models is unclear up to now. 

In this paper, we are trying to address the above issues. To be specific, following the usual prescription of the transfer-matrix-based methods, first we  express the partition function of a 3D classical model in terms of some special two-dimensional transfer matrices, and reduce the problem to the dominant eigenvalue problem of the matrices. Then we solve the dominant eigenvalue problem by representing the dominant eigenvector as a special tensor-network state, namely the projected entangled simplex state (PESS) \cite{PESS2014}, which is efficiently determined through a power iteration procedure analogous to imaginary time evolution, as done for two-dimensional classical models similarly \cite{GV-NUTEBD}. Furthermore, a simple nesting technique is proposed in which the PESS representations of the left and right dominant eigenvectors are designed appropriately so that their inner-product can be expressed as a tensor network with bond dimension $D$ instead of $D^2$, and thus the contraction can be carried out much more efficiently. Combining this nesting technique with the corner transfer matrix renormalization group (CTMRG) algorithm \cite{CTMRG1996, CTMRG2009, tJ2014}, we are able to push the bond dimension $D$ to 20 in this work. For the 3D Ising model, it shows that even if the tensor network state is renormalized by the so-called simple update technique \cite{SU1D2007, SU2D2008, PESS2014} after each evolution step, the obtained local quantities, such as energy density and spontaneous magnetization, are in good consistent with the previous studies, and the estimated critical temperature $T_c$ is about $4.50984(2)$ which has only a relative deviation of about $10^{-4}$ from the best Monte Carlo estimations.

The rest of the paper is organized as follows. In Sec.~\ref{Sec:method}, we introduce some details of the algorithm employed in this paper, including the nesting technique. The numerical results for statistical averages, such as energy density $E$ and spontaneous magnetization $M$, as well as the convergence analysis, are presented in Sec.~\ref{Sec:result}. In Sec.~\ref{Sec:summary}, we summarize our paper, and discuss the possible improvement and promising extensions briefly.

\section{Method}
\label{Sec:method}
\subsection{Tensor-network representation of the partition function}
For concreteness, hereinafter, we will focus on the 3D Ising model. The partition function can be written as
\begin{equation}
	Z = \sum_{\{s\}}\prod_{\langle ij \rangle}e^{\beta s_is_j}, \label{eq:PF}
\end{equation}
where $s_i=\pm 1$ is the spin variable located on the $i$th lattice site, $\beta$ is the inverse temperature, $\langle ij \rangle$ means the product is performed over all the nearest-neighboring bonds, and the summation is over all the spin configurations.

\begin{figure}[htbp]
	\includegraphics[width=0.45\textwidth]{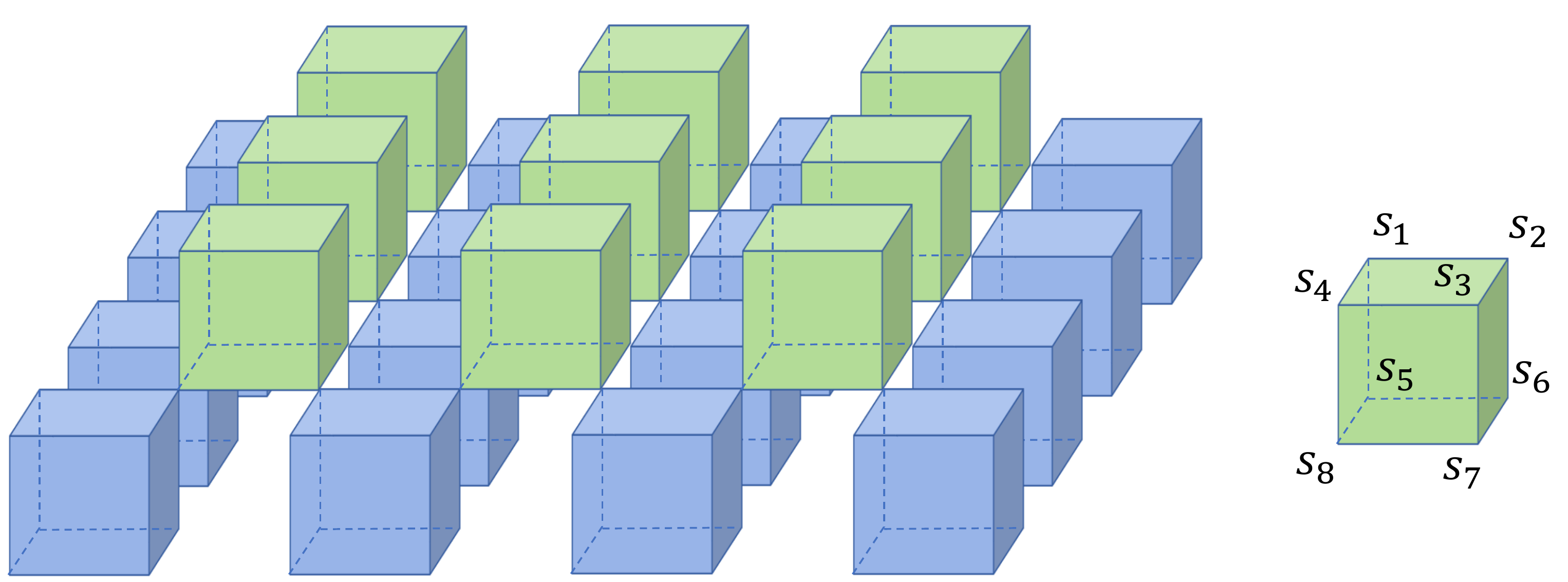}
	\caption{Special transfer matrices in the expression of the partition function of 3D Ising model. (Left) The green and blue cubes constitute transfer matrices $T_2$ and $T_1$, as expressed in Eqs.~(\ref{eq:subtm}) and (\ref{eq:Ztm}). (Right) The definition of local tensor $T$ located in each colored cube, as expressed in Eq.~(\ref{eq:Tdef}). As mentioned in the main text, the vertical direction is referred to as $z$ direction for convenience.}.
	\label{fig:PF}
\end{figure}

For later use, we regroup the product in Eq.~(\ref{eq:PF}) in the unit of cubes that are the building block of a cubic lattice, i.e.,
\begin{equation}
	Z = \sum_{\{s\}}\prod_{\alpha}T^{(\alpha)}, \label{eq:PFCube}
\end{equation}
where $T^{(\alpha)}$ is a rank-8 tensor defined at the $\alpha$th cube. For example, as shown in the right panel of Fig.~\ref{fig:PF}, for a cube where the spin variables residing on the eight vertices are denoted by $s_1$-$s_8$, the local tensor $T$ can be expressed as the following product of twelve Boltzmann weights corresponding to each edge of the cube, respectively, 
\begin{align}
	&T_{s_1s_2s_3s_4s_5s_6s_7s_8} \nonumber \\
	= & \exp \left[\beta (s_1s_2+s_2s_3+s_3s_4+s_4s_1+s_5s_6+s_6s_7+ \nonumber \right.\\
                               	   &\left. s_7s_8+s_8s_5+s_1s_5+s_2s_6+s_3s_7+s_4s_8)\right]. \label{eq:Tdef}
\end{align}

Manifestly, to make Eq.~(\ref{eq:PFCube}) and (\ref{eq:Tdef}) consistent, $T$ should be defined only in two kinds of inequivalent cubes, as denoted as green and blue, respectively, in the left panel of Fig.~\ref{fig:PF}. Thus the two kinds of cubes form an alternative or staggered structure in the vertical direction. This is very similar to the case of the imaginary time evolution in quantum lattice models, where the Trotter-Suzuki decomposition of the evolution operator $e^{-\tau H}$ always leads to an alternative structure in imaginary time $\tau$ direction. In the following, we will use this special structure extensively. For convenience, the vertical direction will be referred to as $z$ direction hereinafter.  

\subsection{Determination of the tensor-network representation of the dominant eigenvector}
\label{lb:SU}
Following the prescription of the transfer-matrix-based method, one needs to express the partition function in terms of some transfer matrices. To this end, firstly, we introduce two tensor-network operators
\begin{align}
	T_1 = \bigotimes_{\alpha\in b} T^{(\alpha)}, \quad T_2 = \bigotimes_{\alpha\in g} T^{(\alpha)}, \label{eq:subtm}
\end{align} 
where the direct products are performed over $T$s defined at {\itshape{blue}} cubes and {\itshape{green}} cubes, respectively, as illustrated in Fig.~\ref{fig:PF}. And then we can identify the following equality:
\begin{equation}
	Z = \Tr\left(T_2T_1\right)^{2n}, \label{eq:Ztm}
\end{equation}
where the length in the z direction is denoted as $2n$ for convenience. It is worth noting that $T_1$ and $T_2$ should be understood as matrices by grouping indices properly in Eqs.~(\ref{eq:subtm}) and (\ref{eq:Ztm}), in order to make the operations for matrices therein meaningful. 

\begin{figure}[htbp]
	\includegraphics[width=0.45\textwidth]{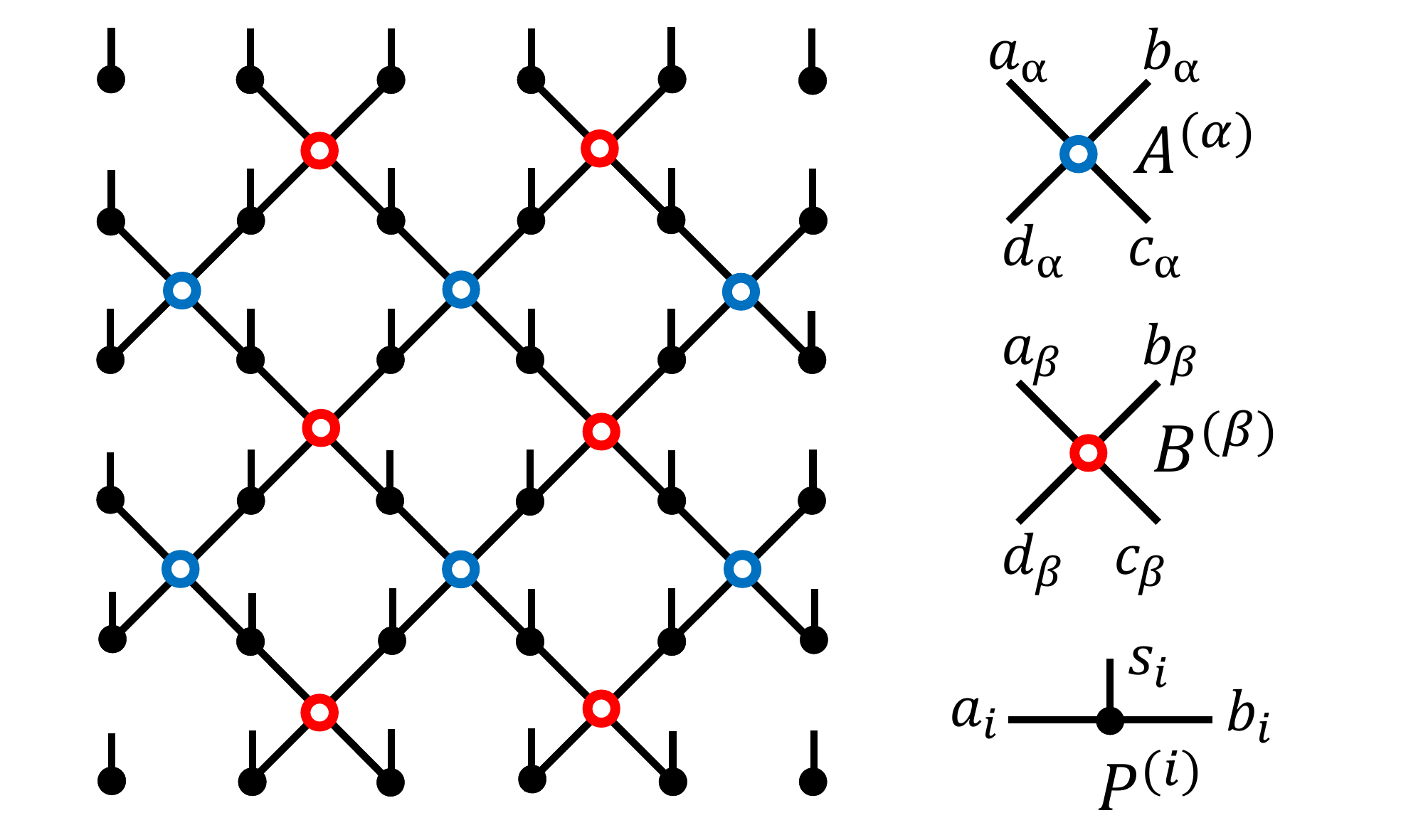}
	\caption{Tensor network representation of the dominant eigenvector of the transfer matrix $T_2T_1$ appeared in Eq.~(\ref{eq:Ztm}), as formulated in Eq.~(\ref{eq:PESS}). The black vertical lines denote the physical configurations $\{s\}$ appeared in Eq.~(\ref{eq:PESS}). In the right panel, the indices of the local tensors are explicitly shown for clarity.}
	\label{fig:PESS}
\end{figure}

Once Eq.~(\ref{eq:Ztm}) is established, the calculation of partition function is immediately reduced to the dominant eigenvalue problem of the transfer matrix $T_2T_1$, which can be solved by the power iteration method. In this paper, we represent the corresponding dominant eigenvector $\ket{\Psi}$ as a PESS form \cite{PESS2014, Schuch-PRB2012}, which can be written as
\begin{equation}
	\ket{\Psi} = \sum_{\{s\}}\mathrm{Tr}(...A^{(\alpha)}_{a_{\alpha}b_{\alpha}c_{\alpha}d_{\alpha}}
	B^{(\beta)}_{a_{\beta}b_{\beta}c_{\beta}d_{\beta}}P^{(i)}_{a_{i}b_{i}}[s_{i}]
	...)|...s_{i}...\rangle,
	\label{eq:PESS}
\end{equation}
as illustrated in Fig.~\ref{fig:PESS}. Here $\alpha$ and $\beta$ denote the coordinates of two inequivalent squares, at the center of which a rank-4 simplex tensor $A$ or $B$ is introduced to characterize the four-spin entanglement in that square. $i$ denotes the coordinates of lattice sites, where a rank-3 projection tensor $P$ is defined with two virtual indices labeled as $a, b...$ and a single physical index labeled as $s$. Every two virtual indices associated with the same bond take the same values. $\mathrm{Tr}$ is over all the repeated virtual indices and $\sum$ is over all the spin configurations $\{s\}$.
In this paper, we employed the $C_4$ rotational symmetry of $A$ and $B$ that corresponds to the symmetry of local tensor $T$ defined in Eq.~(\ref{eq:Tdef}), and the translational symmetry of the 3D lattice, thus there are only one independent $P$, $A$, and $B$ present in Eq.~(\ref{eq:PESS}). More background about the PESS wave function can be found in the Appendix.

\begin{figure}[htbp]
	\includegraphics[width=0.45\textwidth]{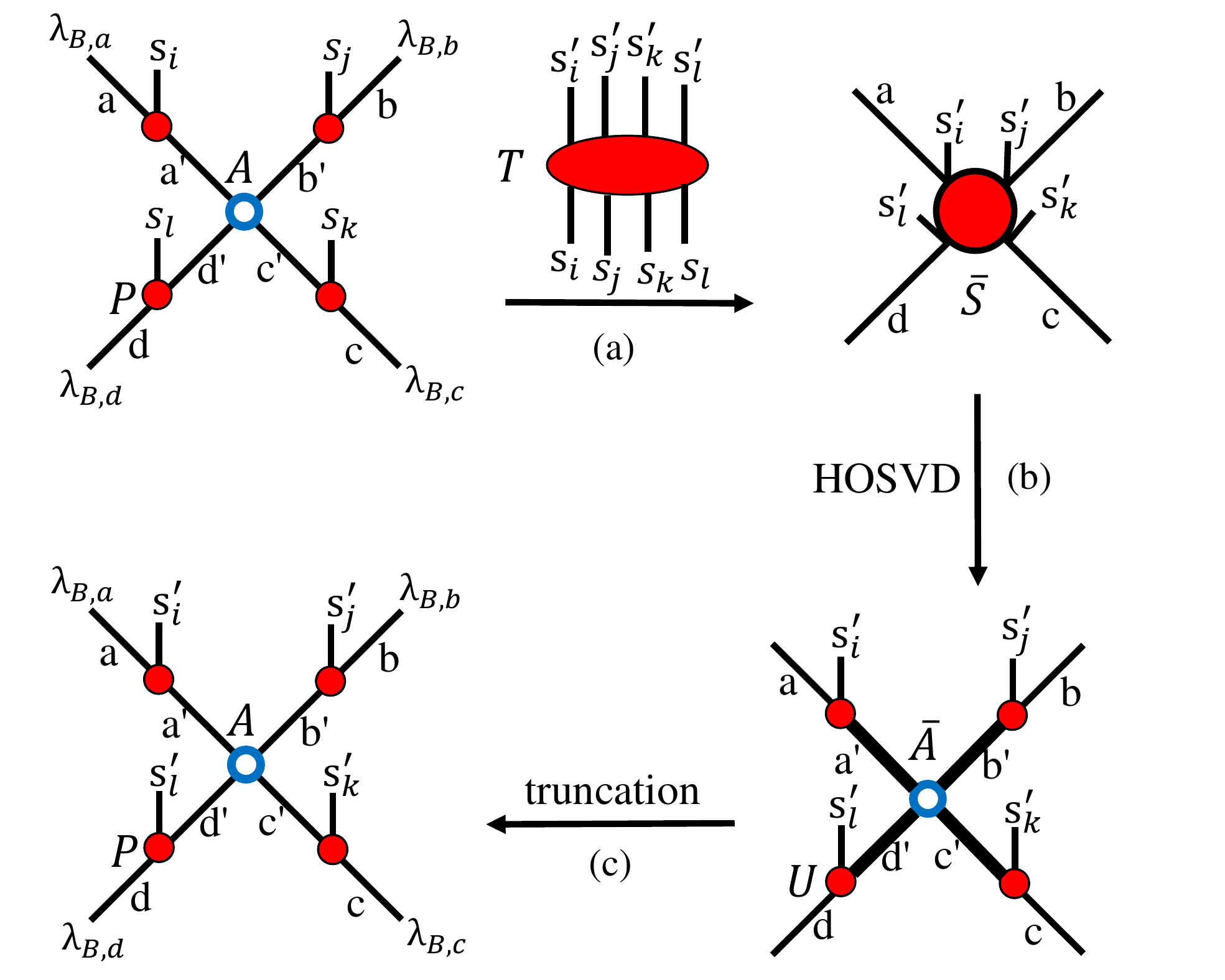}
	\caption{A single projection step of $T$ on the PESS ansatz (take the update of simplex tensor $A$ for example), as described in the main text. To illustrate the operation, the indices of the local tensor $T$ are labeled explicitly. (a) The indices $\{s_is_js_ks_l\}$ in $T$ and $A$ are summed over. A cluster tensor $\bar{S}$ is obtained by further absorbing four mean-field entanglements, i.e., the positive definite vector $\lambda_B$ obtained from the higher-order singular value decomposition (HOSVD) of the $B$ tensor. (b) The HOSVD of $\bar{S}$ is performed to obtain four unitary matrices $U$s and the core tensor $\bar{A}$. (c) Truncate $\bar{A}$ to keep bond dimension $D$, recover the original mean-field entanglement structure, and obtain the updated local tensors $A$ and $P$. For more details, one can refer to the elaboration in Ref.~\cite{PESS2014}.}
	\label{fig:SU}
\end{figure}

As done in time evolution in quantum lattice systems, using the alternative structure of $T_1$ and $T_2$, we can take a similar simple update strategy \cite{SU1D2007, SU2D2008, PESS2014} to determine the variational parameters $A$, $B$, and $P$ in $\ket{\Psi}$. To be specific, starting from a random state $\ket{\Psi_0}$ which has the same structure as $\ket{\Psi}$, we apply $T_1$ and $T_2$ alternatively to $\ket{\Psi_0}$ and update the parameters accordingly by local decompositions of the related clusters after each projection. This procedure is applied repeatedly until the convergence is reached, and the obtained tensor network state provides an approximate representation of $\ket{\Psi}$ \cite{GV-NUTEBD}. Taking $A$ as an example, Fig.~\ref{fig:SU} illustrates how a single projection step is performed. For more details, we suggest referring to Ref.~\cite{PESS2014}. 

\subsection{Inner product between the left and right dominant eigenvectors, and its nesting structure}
\label{Sec:subNTN}
Having described how to obtain the dominant eigenvectors, in this subsection, firstly, we briefly describe how the statistical averages are obtained in the traditional reduced tensor network method (RTN) \cite{OrusReview, NTN-PRB2017}. As long as the dominant eigenvector $\ket{\Psi}$ is obtained, we can use the fundamental formula to calculate the statistical averages of local physical quantities. For example, the average magnetization located at the $i$th site can be determined by
\begin{align}
	M_i = \frac{\sum_{\{s\}}s_i\prod_{\alpha}T^{(\alpha)}}{\sum_{\{s\}} \prod_{\alpha}T^{(\alpha)}} = \frac{\Tr(T_2T_1)^ns_i(T_2T_1)^n}{\Tr(T_2T_1)^n(T_2T_1)^n}, \label{Mdef}
\end{align} 
where we have used the transfer-matrix expression of $Z$ and assumed that $s_i$ sits in the middle in the $z$-direction. In the thermodynamic limit, $n\rightarrow\infty$, we reach
\begin{equation}
	M_i = \frac{\lr{\Psi'|s_i|\Psi}}{\lr{\Psi'|\Psi}}, \quad M = \frac{\sum_{i\in\alpha}M_i}{8}, \label{SponM}
\end{equation}
where $\langle\Psi'|$ is the left dominant eigenvector of $T_2T_1$ and can be derived easily due to the symmetry between $T_1$ and $T_2$. The spontaneous magnetization $M$ is obtained eventually as above by averaging over the eight spins in the same cube due to the translational invariance of the tensor networks. Internal energy $E$ of the bonds can be obtained similarly. To evaluate the ratios like Eq.~(\ref{SponM}), we can follow the widely-used impurity method as discussed in Refs.~\cite{SU2D2008, HHZ-PRB2010, HHZ-PRB2016}, and some details can be found in the Appendix.

\begin{figure}[hbtp]
	\includegraphics[width=0.5\textwidth]{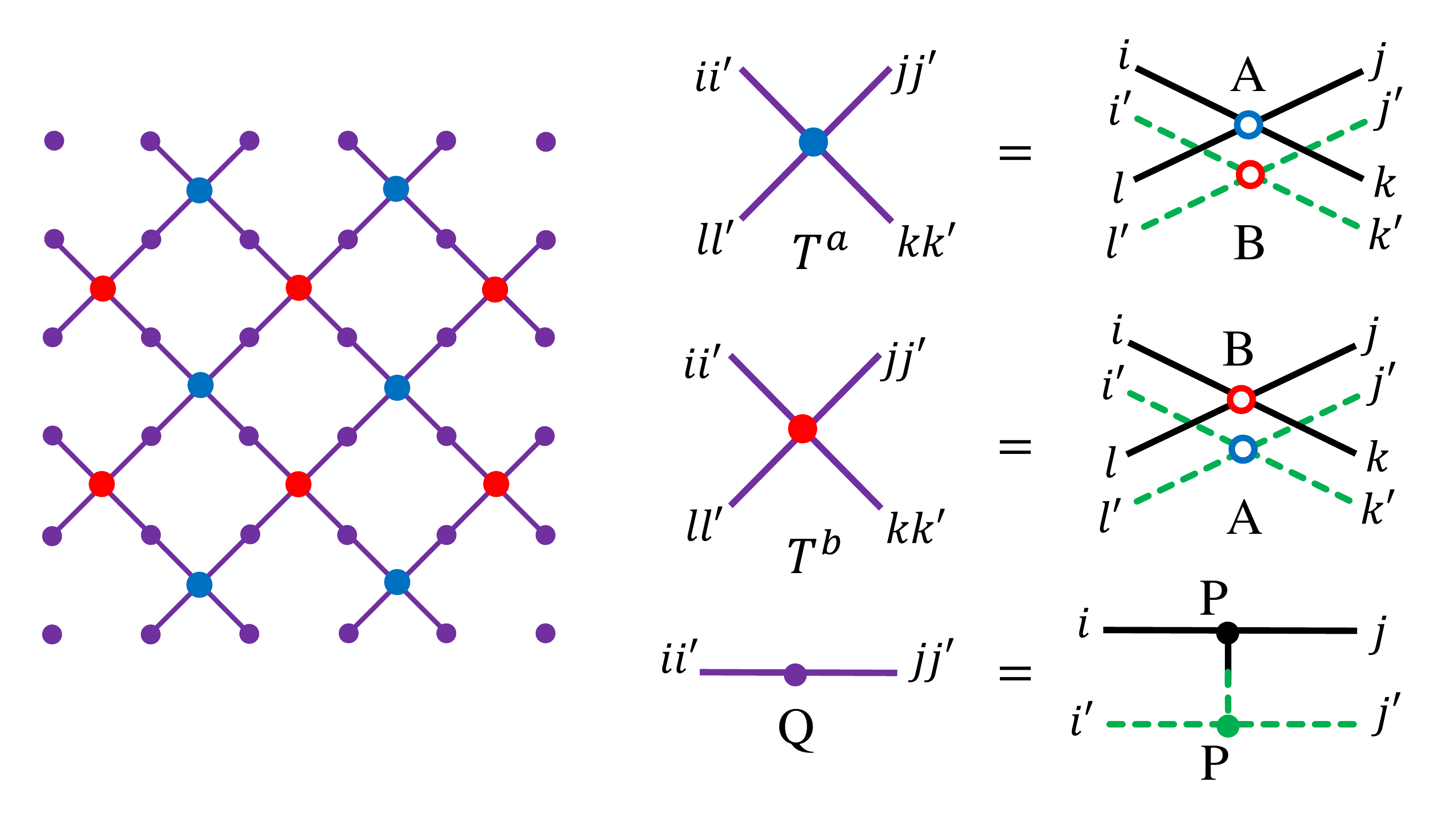}
	\caption{Sketch of $\lr{\Psi'|\Psi}$ appeared in Eq.~(\ref{SponM}). Here $\langle\Psi'|$ is the left dominant eigenvector of $T_2T_1$. The relevance to $\ket{\Psi}$ and thus the structure of the local tensors $T^a$ and $T^b$, come from the symmetry between $T_1$ and $T_2$, i.e., they are transformed to each other by a finite translation in the diagonal direction.}
	\label{fig:RTN}
\end{figure}

As indicated explicitly in Eq.~(\ref{SponM}), the statistical average of a local physical quantity needs to contract a two-dimensional tensor network, which is identical to the expectation value calculation for quantum lattice models in essence. Suppose both $\ket{\Psi}$ and $\bra{\Psi'}$ have bond dimension $D$, then the generated two-dimensional tensor network has dimension $D^2$, as shown in Fig.~\ref{fig:RTN}. When $D$ is small, this tensor network can be contracted by the CTMRG algorithm \cite{CTMRG1996, CTMRG2009, tJ2014} effectively. More introduction of the CTMRG algorithm used in this work is provided in the Appendix.

The simple-update-like algorithm described in Sec.~\ref{lb:SU} is highly efficient, and the leading computational cost scales as $D^5$ in our case. Thus PESS eigenvector with large bond dimensions can be obtained without too much cost. However, contracting a two-dimensional tensor network with bond dimension $D^2$, as shown in Fig.~\ref{fig:RTN}, has a leading cost $D^{12}$ empirically, as discussed in Ref.~\cite{NTN-PRB2017}, and is extremely costly when $D$ is large. A possible solution is the nested tensor network technique, which can reduce the cost to $D^{9}$ \cite{NTN-PRB2017}. Nevertheless, in this paper, we did not take this strategy. Instead, we proposed a much simpler nesting technique by taking advantage of the special cubic lattice structure. This nesting technique can reduce the computational cost similarly as achieved by the nested tensor network method (also scales as $D^9$), keeps the symmetry of the ground state properly, and is especially suitable for cubic systems.

\begin{figure}[hbtp]
	\includegraphics[width=0.4\textwidth]{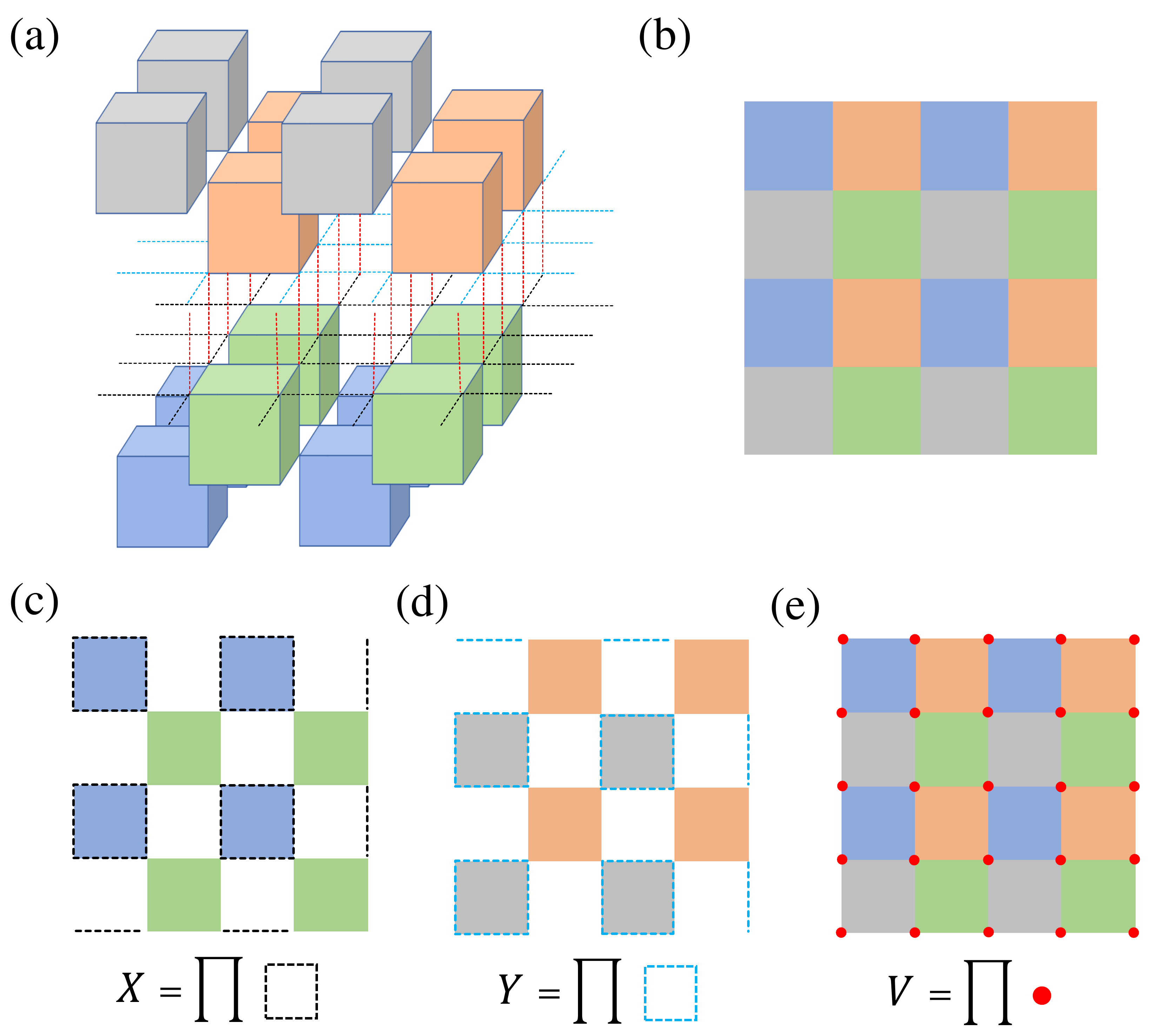}
	\caption{Decomposition of the cubic lattice into three parts, as described in the nesting technique in Sec.~\ref{Sec:method}. (a) The cubes with colors blue, green, orange and grey, constitute the transfer matrices $T_1$, $T_2$, $T_3$, and $T_4$, respectively. The dashed lines with color black, blue, and red, constitute the transfer matrices $X$, $Y$, and $V$, respectively. The full partition function can be represented in terms of these transfer matrices, as expressed in Eq.~(\ref{eq:Znest}). (b) The spatial distribution of the cubes in $xy$ plane, for the sixteen cubes shown in (a). (c) Relative location of X (black dashed lines) to $T_1$ (blue) and $T_2$ (green), and this explains how the operation $X|\Psi\rangle$ is performed. (d) Relative location of Y (blue dashed lines) to $T_3$ (orange) and $T_4$ (grey), and this explains how the operation $\langle\Phi|Y$ is performed. (e) Relative location of V (on-site red dots) to all four $T$s, and this explains how  the operation $\langle\tilde{\Phi}|V|\tilde{\Psi}\rangle$ is performed. Note that (b)--(e) display the projection of (a) onto the horizontal plane, therefore the simplex tensors appeared in the PESS ansatz Eq. (6) and updated in Fig. 3, reside exactly in the center of the colored squares.}
	\label{fig:TM4}
\end{figure}

To see how it works, we first divide the cubic lattice into three parts, namely two bulks whose contribution to the full partition function can be expressed in terms of transfer matrices, and the surface part which connects the two bulks and combines as the whole cubic lattice. This means we have rewritten the partition function in another manner, as illustrated in Fig.~\ref{fig:TM4}(a), 
\begin{equation}
	Z = \Tr (T_4T_3)^nYVX(T_2T_1)^n, \label{eq:Znest}
\end{equation}  
where $T_1$ and $T_2$ are the transfer matrices corresponding to the lower bulk, $T_3$ and $T_4$ correspond to the upper bulk similarly, and $X$, $Y$, $V$ are auxiliary matrices to describe the surface part. Here we intentionally decompose the lower and upper parts in different manners, and their relative locations projected in the $xy$ plane are shown in Fig.~\ref{fig:TM4}(b). Mathematically,
\begin{equation}
	X = \prod_{\lr{ij}_t\not\in g} e^{\beta s_is_j}, \quad Y = \prod_{\lr{ij}_b\not\in o} e^{\beta s_is_j}, \quad V = \prod_{\lr{ij}_{bt}} e^{\beta s_is_j},
	\label{eq:XYZ}
\end{equation}
which are explicitly shown in Figs.~\ref{fig:TM4}(a), and \ref{fig:TM4}(c)--\ref{fig:TM4}(e). Here $X$ and $Y$ can be understood as onsite diagonal matrices, $\lr{ij}_t\not\in g$ means the product is over the nearest spin pairs between green cubes on the top surface of the lower bulk, and $\lr{ij}_b\not\in o$ means the product is over the nearest spin pairs between orange cubes at the bottom surface of the upper bulk. $\lr{ij}_{bt}$ means $V$ is the product of Boltzmann weights corresponding to all the bonds connecting the two bulks in the $z$ direction. In Fig.~\ref{fig:TM4}, the dashed bonds corresponding to $X$, $Y$, and $V$, are denoted as black, blue, and red, respectively. Note in Eq.~(\ref{eq:Znest}), the size in the $z$ direction is assumed as $2n$+1 for convenience. 

In the thermodynamic limit, the evaluation of Eq.~(\ref{eq:Znest}) is again reduced to the dominant eigenvalue problem as before, and the statistical averages can be determined similarly. For example, the bond energy in the $z$ direction can be obtained by 
\begin{equation}
	E_{p} = \frac{\lr{\Phi|YV'X|\Psi}}{\lr{\Phi|YVX|\Psi}}, \quad V' = (-s_as_be^{\beta s_as_b})\prod_{\lr{ij}_{bt}\neq p}e^{\beta s_is_j},
	\tred{\label{eq:t1}}
\end{equation}
where $\langle\Phi|$ is the left dominant eigenvector of $T_4T_3$, and $V'$ differs from $V$ by only a single vertical bond denoted as $p$ with $s_a$ and $s_b$ located at its ends. Finally we have
\begin{equation}
	E_p = \frac{\lr{\tilde{\Phi}|V'|\tilde{\Psi}}}{\lr{\tilde{\Phi}|V|\tilde{\Psi}}}, \quad \langle\tilde{\Phi}|\equiv\langle\Phi|Y, \quad  \ket{\tilde{\Psi}}\equiv X\ket{\Psi},
	\label{eq:Erep}
\end{equation}

\begin{figure}[hbtp]
	\includegraphics[width=0.45\textwidth]{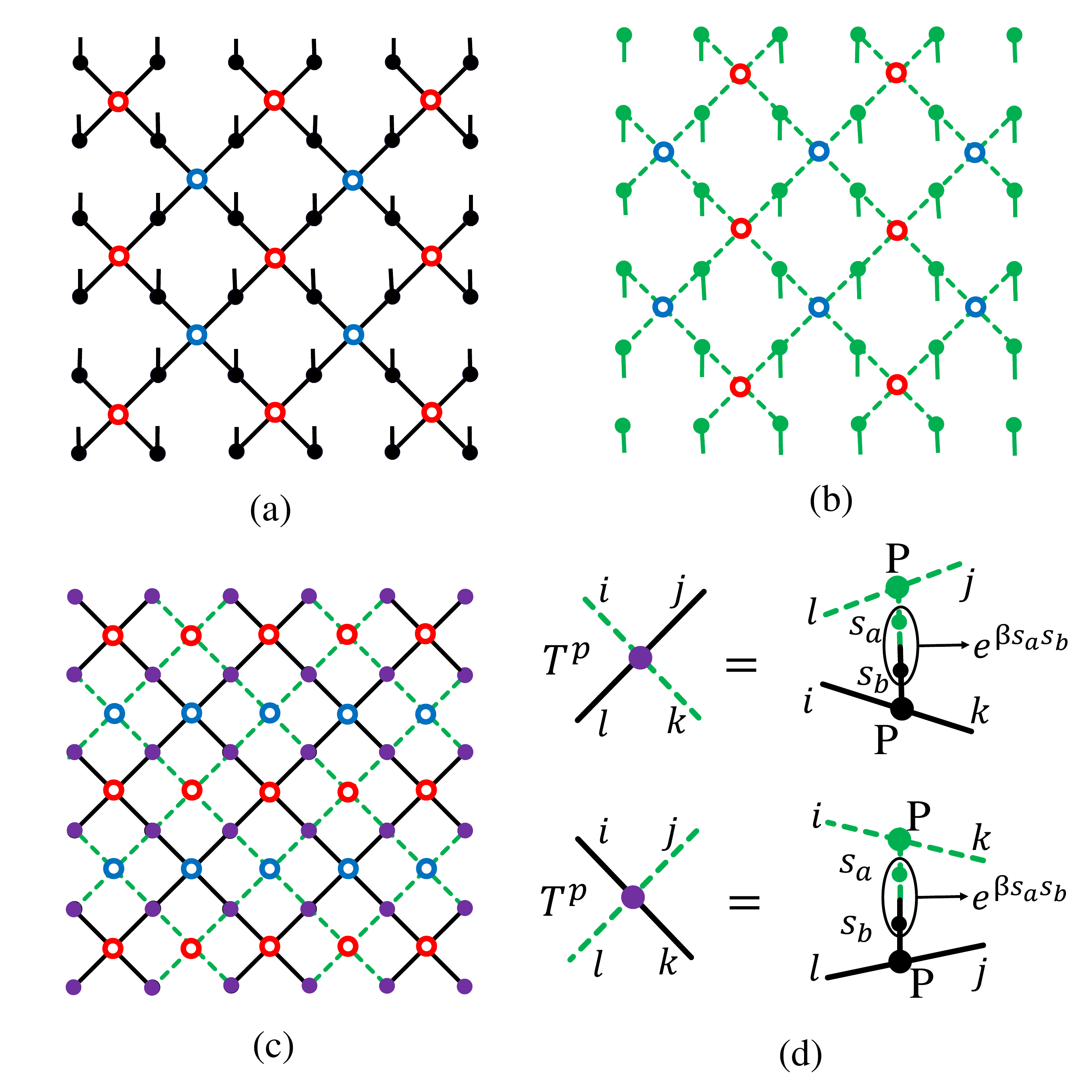}
	\caption{Illustration of the nesting structure of $\lr{\tilde{\Phi}|V|\tilde{\Psi}}$ appeared in Eq.~(\ref{eq:Erep}). The spatial location corresponds to Fig. 5 exactly. (a) Structure of the ket state $\ket{\tilde{\Psi}}$, whose virtual bonds are indicated as solid black lines. (b) Structure of the bra state $\bra{\tilde{\Phi}}$, whose virtual bonds are indicated as dashed green lines. (c) Structure of the inner product $\lr{\tilde{\Phi}|V|\tilde{\Psi}}$, where new rank-4 tensor $T^p$ is defined in Eq.~(\ref{eq:Tp}) and illustrated in (d). Here we have used the symmetry between the two wavefunctions, and both  definitions of $T^p$ hold, although it is unnecessary for the nesting method. The local tensors are the parameters in $\ket{\tilde{\Psi}}$ in this context, not to be confused with parameters in $\ket{\Psi}$ in Fig.~\ref{fig:PESS}. Clearly, the bond dimension of the resulting tensor network is not squared, which is different from the case in Fig.~\ref{fig:RTN}.}
	\label{fig:NTN}
\end{figure}

The key observation is that $\langle\tilde{\Phi}|$ and $\ket{\tilde{\Psi}}$ have a similar structure , but the parameters are distributed separately in space, i.e., their simplex tensors are defined separately in space at different sets of squares, which can be seen from Figs.~\ref{fig:TM4}(b)--\ref{fig:TM4}(d) and Figs.~\ref{fig:NTN}(a)--\ref{fig:NTN}(c) straightforwardly. And this leads to a great advantage, i.e., if the wavefunctions $\langle\tilde{\Phi}|$ and $\ket{\tilde{\Psi}}$ are approximated by PESS with bond dimension $D$, then the generated two-dimensional tensor network also has bond dimension $D$, instead of $D^2$. To be specific, let us take the denominator as an example. Exploring the symmetry between $\langle\tilde{\Phi}|$ and $\ket{\tilde{\Psi}}$, the resulting tensor network 
are composed of three parameters, i.e., the local tensors in $\ket{\tilde{\Psi}}$ and $\ket{\tilde{\Phi}}$, and a new local tensor $T^p$ derived from $P$ therein, i.e., 
\begin{eqnarray}
	T^p_{ijkl} = \sum_{s_as_b}P_{lj}[s_a]P_{ki}[s_b]e^{\beta s_as_b}, \label{eq:Tp}
\end{eqnarray}
as illustrated in Figs.~\ref{fig:NTN}(c)--\ref{fig:NTN}(d). Therefore, compared with the reduced method, which generates a tensor network with a squared bond dimension, this nesting technique can greatly reduce the cost and leads to much more efficient contraction, as achieved in Ref.~\cite{NTN-PRB2017} similarly. Considering the fact that the environment dimension $\chi$ used in the CTMRG iterations has the order $D^2$ empirically, the nesting technique can reduce the computational cost from $D^{12}$ to $D^9$. In this paper, we push $D$ to 20 with the help of this technique.

\section{Results}
\label{Sec:result}
In this paper, we focus on the 3D Ising model, which is of long-standing interest in statistical physics and condensed matter physics.
Although there is no analytical solution as in two-dimensional case, the higher-order tensor renormalization group (HOTRG) \cite{HOTRG-PRB2012, SW-CPL2014} and Monte Carlo methods \cite{Blote-JPA1996, Blote-PRE, Hasen-PRB2010, Landau-PRE2018} have provided very accurate numerical data, which confirms a second-order finite temperature phase transition. Therefore, this model provides a suitable touchstone to test new numerical algorithms in higher dimensions.

In this paper, as described in Sec.~\ref{Sec:method}, we use the developed evolution method combined with the simple update technique to determine the tensor-network representation of the dominant eigenvectors of the transfer matrices. Furthermore, to push $D$ to a larger value (up to 20 in this paper), in the expectation value calculations, we always employ the nesting technique described in Sec.~\ref{Sec:subNTN} to improve the efficiency of CTMRG algorithm. Note that when $D$ is small, and the conventional reduced method can be efficiently performed, the results obtained by the CTMRG with and without nesting technique are consistent, as illustrated and discussed in the Appendix. 

\begin{figure}[htbp]
	\includegraphics[width=0.5\textwidth]{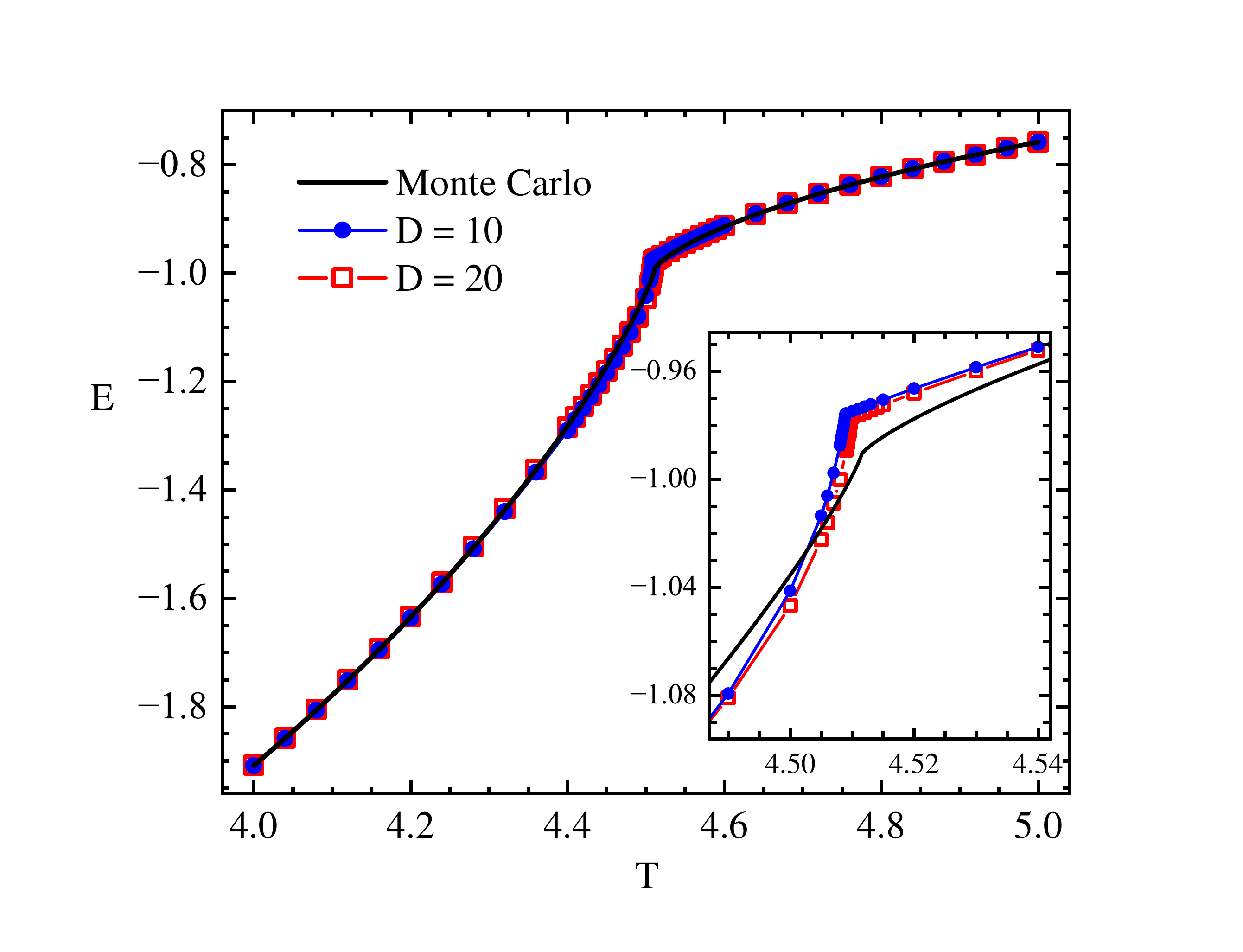}
	\caption{Energy estimation obtained from the simple update and the nesting method. The Monte Carlo data from Ref.~\cite{Blote-PRE} are plotted as a reference.}
	\label{fig:Energy}
\end{figure}

\begin{figure}[htbp]
	\includegraphics[width=0.5\textwidth]{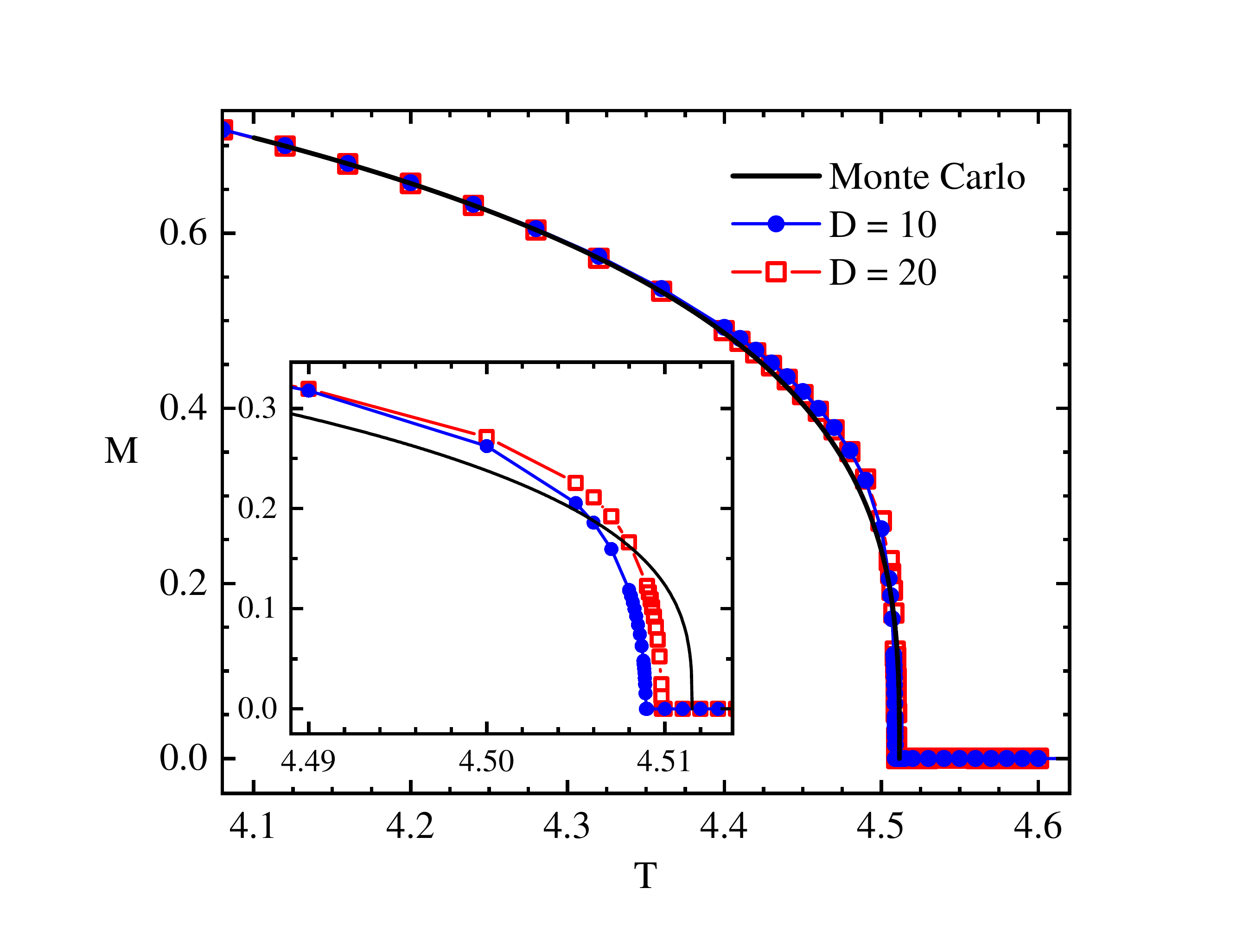}
	\caption{Magnetization estimation obtained from the simple update and the nesting method. The Monte Carlo data from Ref.~\cite{Blote-JPA1996} are plotted as a reference.}
	\label{fig:Mag}
\end{figure}

The result of internal energy $E$ is shown in Fig.~\ref{fig:Energy}. The reference curve denotes high-precision Monte Carlo data obtained in Ref.~\cite{Blote-PRE}. It shows that the proposed method can give consistent results at the off-critical region, while near the critical point, enlarging the bond dimension $D$ can produce more accurate results as expected. The spontaneous magnetization $M$ shows similar behavior, as shown in Fig.~\ref{fig:Mag}, where the reference curve is obtained from Monte Carlo \cite{Blote-JPA1996}. For both quantities, our result coincides well with previous studies, and the singular behavior can be seen clearly. 

\begin{figure}[htbp]
	\includegraphics[width=0.5\textwidth]{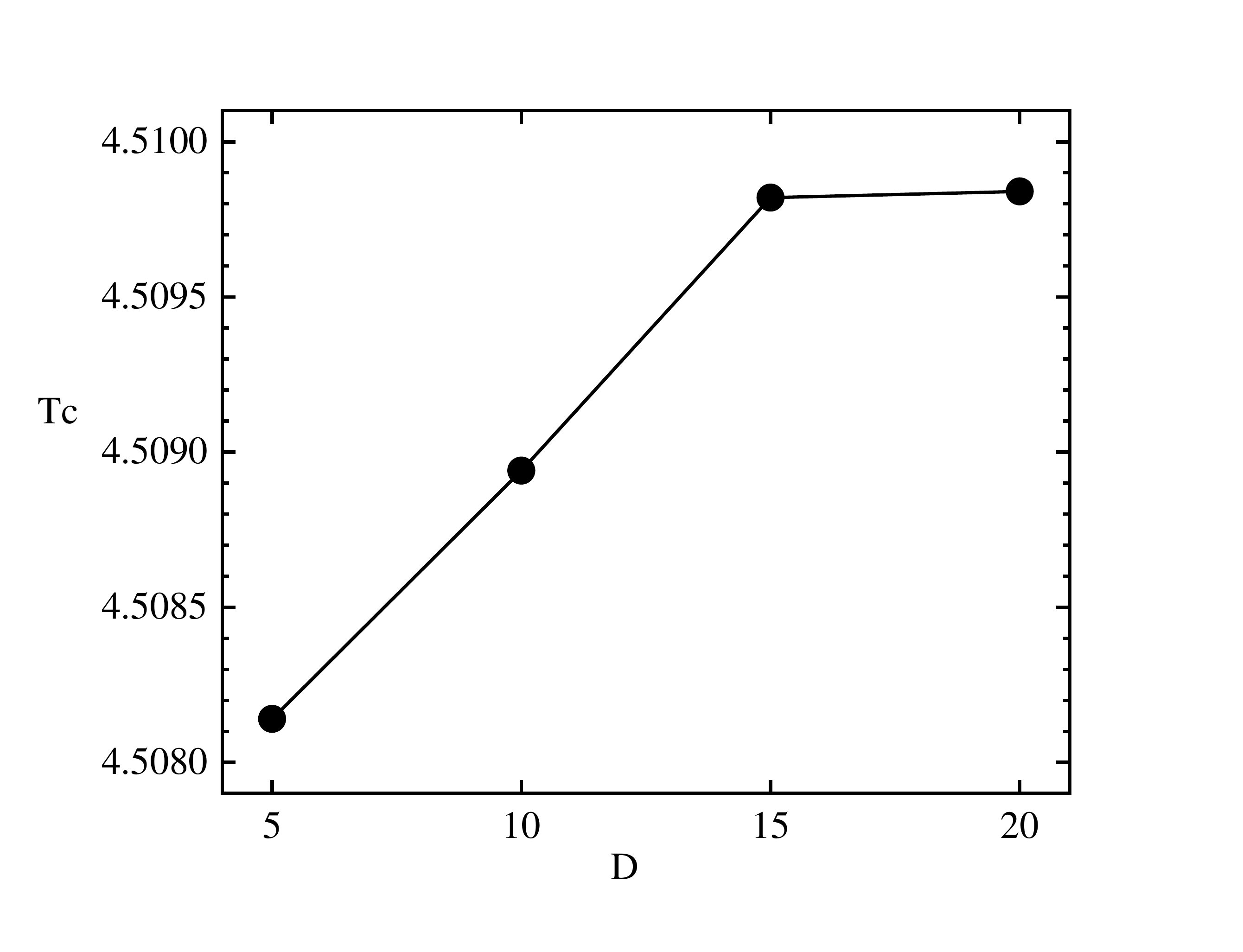}
	\caption{Critical temperature obtained from the magnetization $M$, with respect to the bond dimension $D$ of the tensor network state representation of the transfer matrices' dominant eigenvectors. The critical temperature is estimated as the lowest temperature which has almost zero magnetization, as shown in Fig.~\ref{fig:Mag}.}
	\label{fig:Tc}
\end{figure}

From $E$ and $M$, and the critical temperatures derived as shown in Fig.~\ref{fig:Tc}, it seems that the proposed method tends to underestimate the critical temperature a little, which is different from the coarse-graining tensor renormalization group methods as shown, e.g., in Refs.~\cite{SRG2009, HOTRG-PRB2012}. This is probably related to the simple update technique used in the evolution process, since it essentially provides a Bethe lattice approximation of the dominant eigenvector \cite{WL-PRB2012, HJL-PRB2016} and thus tends to be disordered at finite temperature, especially at the temperature close to but lower than critical point where the approximation cannot give a good estimate of the correlation length. When $D=20$, the estimated critical temperature $T_c$ is about $4.50984(2)$, which has only about $10^{-4}$ relative deviation from the best estimation $T_c\sim 4.51152$ \cite{HOTRG-PRB2012, SW-CPL2014, Landau-PRE2018}. The comparison with other variational calculations is summarized in Table~\ref{tab:Tc}. Although obtaining a benchmark result is not our motivation, it shows that our method performs rather well. Compared to the most recent PEPS calculations \cite{FV-PRE2018, LV-PRL2022}, though our result seems slightly less accurate, our method does not need other auxiliary techniques, such as ignoring the long tails of magnetization with careful fitting \cite{FV-PRE2018}, scaling hypothesis and data collapse \cite{LV-PRL2022}, bootstrap exponent fixing \cite{LV-PRL2022}, etc. The accuracy of the critical temperature and the critical exponent can be further improved by further refining the simple-updated wave functions through variational calculations  and performing these advanced techniques. 

\begin{table}
\begin{tabular}{c|c}
	\hline \hline 
	 Method & $T_c$  \\
	\hline 	
	Monte Carlo \cite{Landau-PRE2018}       &  4.51152326(11)     \\
	HOTRG \cite{HOTRG-PRB2012, SW-CPL2014}    & 4.51152469(1) \\
	KWA-based TNS$^{\ast}$ \cite{Nishino-KWA}  &  4.5788 \\
	Vertex-type TNS$^{\ast}$ \cite{Nishino-TPVA}  & 4.5392 \\ 
	TPVA$^{\ast}$ \cite{Nishino-KWA, TPVA-2}  & 4.5704, 4.554 \\
	TNS$^{\ast}$  \cite{Nishino-TPVA}  & 4.504 \\
	Algebraic variation$^{\ast}$ \cite{SGC-PLA2006} & 4.547 \\
	TNS data fitting$^{\ast}$ \cite{FV-PRE2018}       &  4.5118057(41) \\	
	TNS data collapse$^{\ast}$ \cite{LV-PRL2022} & 4.5104, 4.51170 \\	
	Simple update (this paper)         &  4.50984(2)           \\
	\hline \hline
\end{tabular}
\caption{Comparison of the critical point $T_c$ for the 3D Ising model obtained from different tensor-network-state-based methods. The Monte Carlo and HOTRG results are included as references. Note that  for the methods denoted by $(^\ast)$, variational calculations have been employed and the bond dimensions are rather small ($D\sim 4$). Our results are derived from the magnetization data with $D=20$ simple-updated wave function.}
\label{tab:Tc}
\end{table}

\begin{figure}[htbp]
	\includegraphics[width=0.5\textwidth]{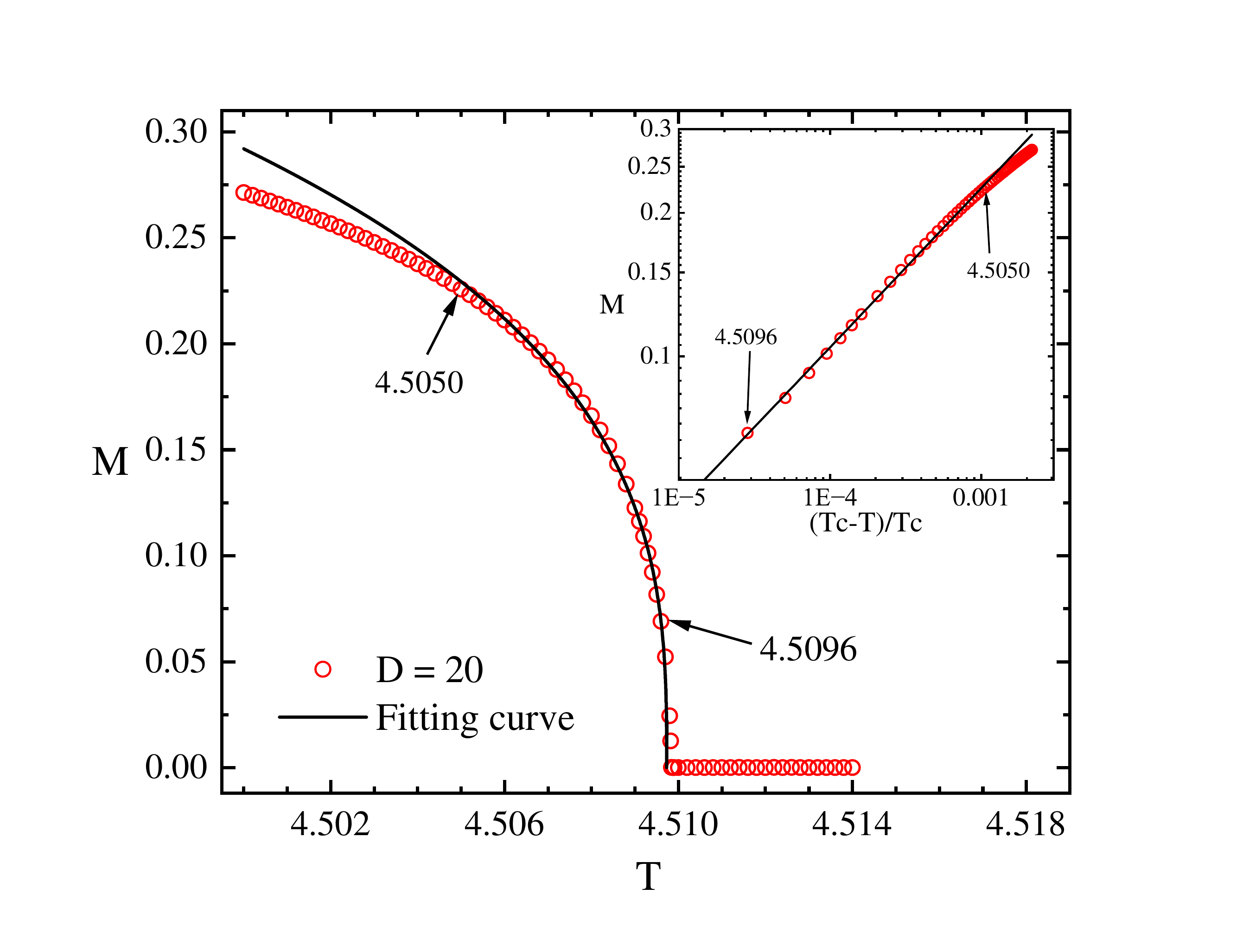}
	\caption{Critical exponent $\beta$ for the magnetization obtained from $D = 20$ simple-updated wave function and $\chi = 120$ nesting CTMRG method. $\beta$ is estimated as 0.335(16), and the data obtained at temperature ranging from 4.5050 to 4.5096 are used for fitting.}
	\label{fig:Beta}
\end{figure}

Furthermore, as shown in Fig.~\ref{fig:Beta}, the critical exponent can be obtained by fitting the magnetization data near critical temperature according to the scaling relation 
	\begin{eqnarray}
		M \sim \left(1 - T/T_c\right)^{\beta}.
	\end{eqnarray}
	We find that the exponent $\beta=0.335(16)$, close to the Monte Carlo \cite{Hasen-IJMPC2001} (0.3262), conformal bootstrap \cite{Vichi-JHEP2016} (0.326419), and HOTRG \cite{HOTRG-PRB2012} (0.3295) results. 

\begin{figure}[htbp]
	\includegraphics[width=0.5\textwidth]{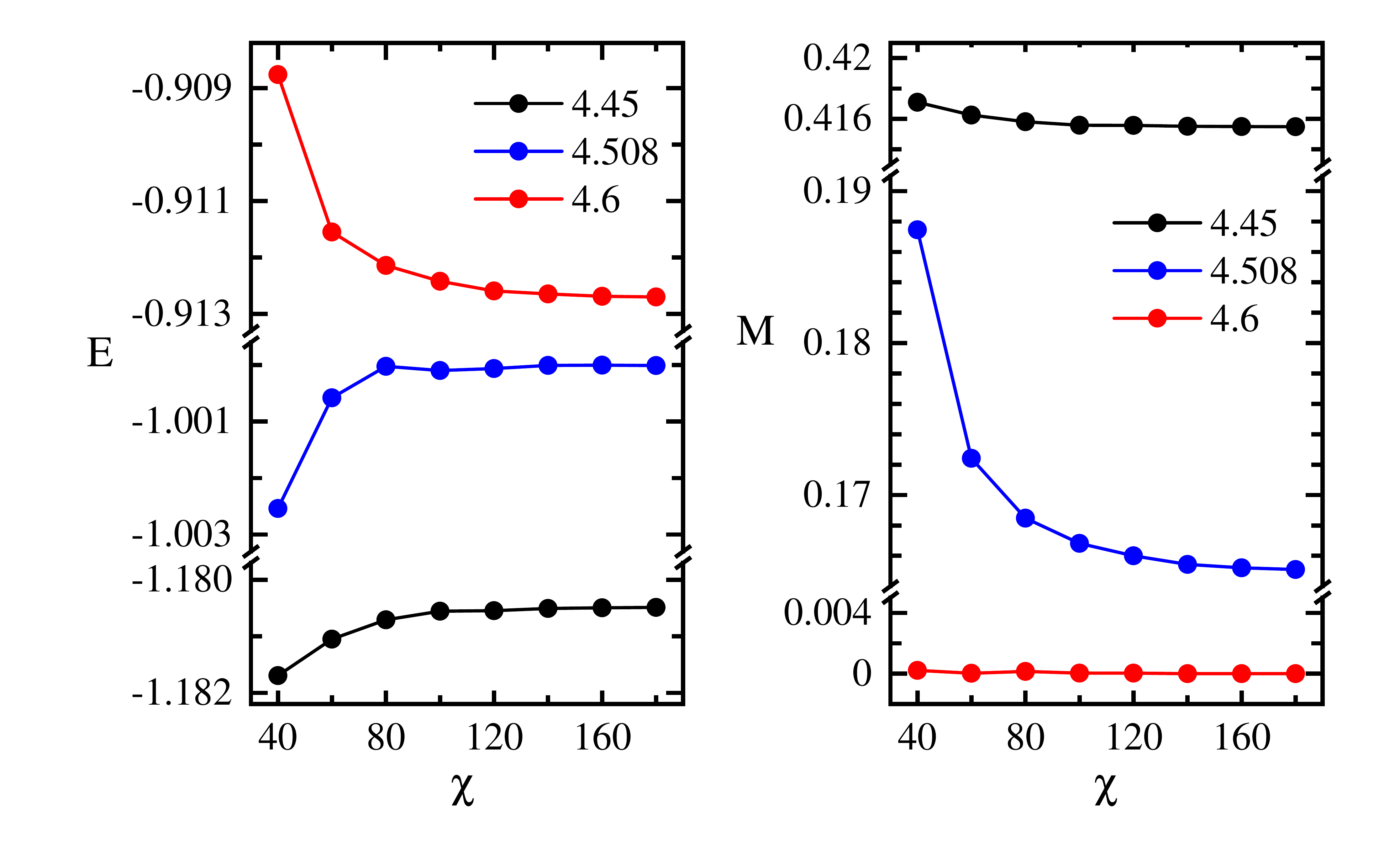}
	\caption{Convergence analysis about energy and magnetization for $D=20$. Three typical temperatures are chosen, i.e., temperature in symmetry-breaking phase (4.45), close to critical point (4.508), and in disordered phase (4.6).}
	\label{fig:Chi}
\end{figure}

As to the nesting technique proposed in this paper, its convergence with respect to the bond dimension $\chi$ of the environment tensors used in CTMRG is shown in Fig.~\ref{fig:Chi}, for $D=20$. It shows clearly that for all the three typical temperatures ranging from symmetry-breaking phase to paramagnetic phase, the convergence is at least equally satisfying, compared with the nested tensor network method proposed in Ref.~\cite{NTN-PRB2017}. Particularly, even at the temperature that is very close to the critical value, the convergence is already very acceptable when $\chi=180$ for $D=20$, with an error about $10^{-6}$ for $E$ and $10^{-4}$ for $M$. This is a very nice feature for tensor-network methods, especially for large $D$, as discussed in Ref.~\cite{NTN-PRB2017} in detail.   

\section{Summary}
\label{Sec:summary}
In this paper, we propose an efficient numerical method to contract the 3D tensor networks with translational invariance. The result of the contraction is expressed in terms of some transfer matrices, whose dominant eigenvectors are represented as the PESS form and determined by power iterations analogous to the imaginary time evolution algorithm \cite{GV-TEBD, GV-NUTEBD}. Especially, the PESS representations of the left and right dominant eigenvectors are designed appropriately so that their innerproduct has a nesting structure and can be expressed as a tensor network with bond dimension $D$ instead of $D^2$, and thus the contraction can be carried out much more efficiently. As to the 3D Ising model, it can give very consistent results for both energy and magnetization with previous studies, even though only the simple update strategy is employed to update the tensor network states. The convergency of the nesting technique is shown to be at least equally good to that of the nested tensor network method \cite{NTN-PRB2017}. When $D=20$, we obtain a critical temperature about 4.50984(2), which is very close to the best-known estimation. As far as we know, this is probably the successful trial of applying the imaginary-time-evlution-like method to 3D classical models without variational update procedure, thus this extends the application scope of tensor network states. The application to other interesting but unsolvable classical models in three dimensions, such as the Potts model \cite{SW-CPL2014}, clock model \cite{Sandvik-PRL2020}, dimer model \cite{FV-PRE2018}, and lattice gauge models \cite{LQCD2013}, is straightforward. 

As mentioned in Sec.~\ref{Sec:result}, the simple update strategy employed in this paper is probably a reason why the obtained $T_c$ tends to be underestimated. Besides the scaling hypothesis and data collapse technique \cite{LV-PRL2022}, this tendency could be eased in two different manners. One is to resort to the more involved but more accurate update methods, such as the 
cluster update \cite{CU-PRB2016} and the full update strategies \cite{FU-PRL2008, FU-PRB2014}, which consider the renormalization effect of the environment better. The other possible way is to utilize more sophisticated ansatz and evolution techniques, such as PESS ansatz with stronger simplex entanglement \cite{PESS2014}, the recently proposed regularized scheme of time evolution \cite{LXC-arXiv2022}, and the so-called minimal canonical form of tensor network states \cite{AA-arXiv2022}, all of which are also expected to produce more accurate representation of the dominant eigenvectors. In both cases, the nesting technique proposed in this paper can still be applied without changing too much, and the improvement in performance can be expected.
 
At last, it is worth mentioning that the proposed nesting technique in this paper can also be extended to the two-dimensional quantum lattice models. In order to achieve this, one needs to represent the partition function in a different manner, e.g., similarly as we have done in Eq.~(\ref{eq:Znest}), so that two representations of the ground state with separated distributions in space can be used simultaneously, e.g., as illustrated in Fig.~\ref{fig:TM4}(b) and expressed in Eq.~(\ref{eq:Erep}). This can be advantageous in the study of models which have square structure and separatable block interactions, such as Shastry-Sutherland model \cite{SSL-From1981, Corboz-PRL2014}, checkboard system \cite{HYZ-arXiv2020}, and so on. Exploring its potential and limits is an interesting and promising project, and we would like to leave it as a future pursuit. 

\section*{Acknowledgement}
We thank Zhong-Cao Wei and Zhi-Yuan Liu for their contribution in the early stage of this work. We are supported by the National R$\&$D Program of China (Grants No. 2017YFA0302900 and No. 2016YFA0300503), the National Natural Science Foundation of China (Grants No. 12274458, No. 11888101, No. 11874095, and No. 11774420), and by the Research Funds of Renmin University of China (Grant No. 20XNLG19). L.-P.Y. and Y.F.F contribute equally to this work.

\section*{Appendix:FURTHER BACKGROUND AND COMPARISONS}
In this Appendix, some further background on the concepts and techniques involved in this paper, and some relevant comparisons are provided.

\subsection*{1. PESS wave function ansatz}
The PESS ansatz is particularly important in the nesting technique developed in this paper. It was proposed in Ref.~\cite{PESS2014}, and can be regarded as a generalization of the projected entangled pair state (PEPS) ansatz proposed in Ref.~\cite{PEPS2004}. It extends pair entanglement well characterized by PEPS to simplex entanglement that might be more important in some situations, and has been successfully applied to a series of frustrated spin systems, such as kagome spin liquid systems \cite{Liao-PRL2017, YJK-PRB2018}, chiral spin liquid systems \cite{RW-PBL2022, Sen-PRB2022}, quantum spin-orbital liquid systems \cite{LQ-PRB2022, LY-PRB2023}, etc. 

As a wavefunction ansatz, the defining feature of a PESS is the existence of simplex tensors in it, each of which connects more than two physical degrees of freedom and captures the many-body entanglement among them. For example, in Eq.(\ref{eq:PESS}) and Fig.~\ref{fig:PESS}, simplex tensors $A^{\alpha}$ and $B^{\beta}$ are introduced to characterize the entanglement among the four spins defined on the vertices of the $\alpha$th and $\beta$th squares respectively. For any given base ket $|\{s\}\rangle$, the corresponding coefficient in the superposition is given by the $\mathrm{Tr}$ operation, which means the contraction of a two-dimensional tensor network and can be accomplished by, e.g., the CTMRG method. For general discussions about tensor-network state, Refs.~\cite{PEPS2004, SimBook2018, OrusReview} are excellent references covering almost all important topics.

In this paper, the translational symmetry and the $C_4$ rotational symmetry of the simplex tensors are employed, thus we have only one independent $A$, $B$, and $P$ as variational parameters that need to be determined by the simple-update-aided imaginary-time evolution method, as discussed in the main text and Ref.~\cite{PESS2014}.
\subsection*{2. CTMRG algorithm for arbitrary supercell} 
The CTMRG is a highly-efficient and extensively-used algorithm to contract two-dimensional tensor networks, and in this paper it has been used to calculate the expectation values, since the numerator and denominator in both Eqs.~(\ref{SponM}) and (\ref{eq:Erep}), as well as other terms similar to the inner product of two tensor-network states, can be represented as two-dimensional tensor networks exactly. In this paper, for a given tensor-network state with bond dimension $D$, the difference in the expectation value calculation between the contraction of a normal network and a network with nesting structure, lies in the fact that the normal one has bond dimension $D^2$ while the nested one has dimension $D$. Therefore, though both the two kinds of tensor networks can be contracted by CTMRG algorithm, the leading cost differs greatly, i.e., $D^{12}$ for the normal one and $D^9$ for the nested one (can be reduced to $D^{10}$ and $D^{8}$ respectively by some advanced techniques, e.g., partial SVD.)

In this paper, we used the CTMRG algorithm for arbitrary supercell that was proposed in Ref.~\cite{tJ2014}. For the normal network depicted in Fig.\ref{fig:RTN}, the supercell size is $2\times 2$ (one can always absorb the Q tensors into $T^a$ and/or $T^b$), while for the nested network depicted in Fig.~\ref{fig:NTN}, the supercell size is $4\times 4$. In the CTMRG algorithm, each independent tensor in the supercell is associated with four corner tensors and four edge tensors, which mimic its effective environment and are used to extract the expectation values. For a supercell with size $m\times n$, we have stored $4mn$ corners and $4mn$ edges in total, which are initialized as random numbers and are updated iteratively until convergence is reached. The detailed iterative relations among these corners and edges are determined according to their relative spatial locations, and have been given clearly in Ref.~\cite{tJ2014}. For general discussions about the CTMRG algorithm, Ref.~\cite{OrusReview} provide excellent and friendly reviews. For the most recent development, Refs.~\cite{XFL-CPL2022, Lukin-PRB2023} provide promising trials.
\subsection*{3. Impurity method in tensor-network states}
\begin{figure}[htbp]
\vspace{0.6cm}
	\includegraphics[width=0.48\textwidth]{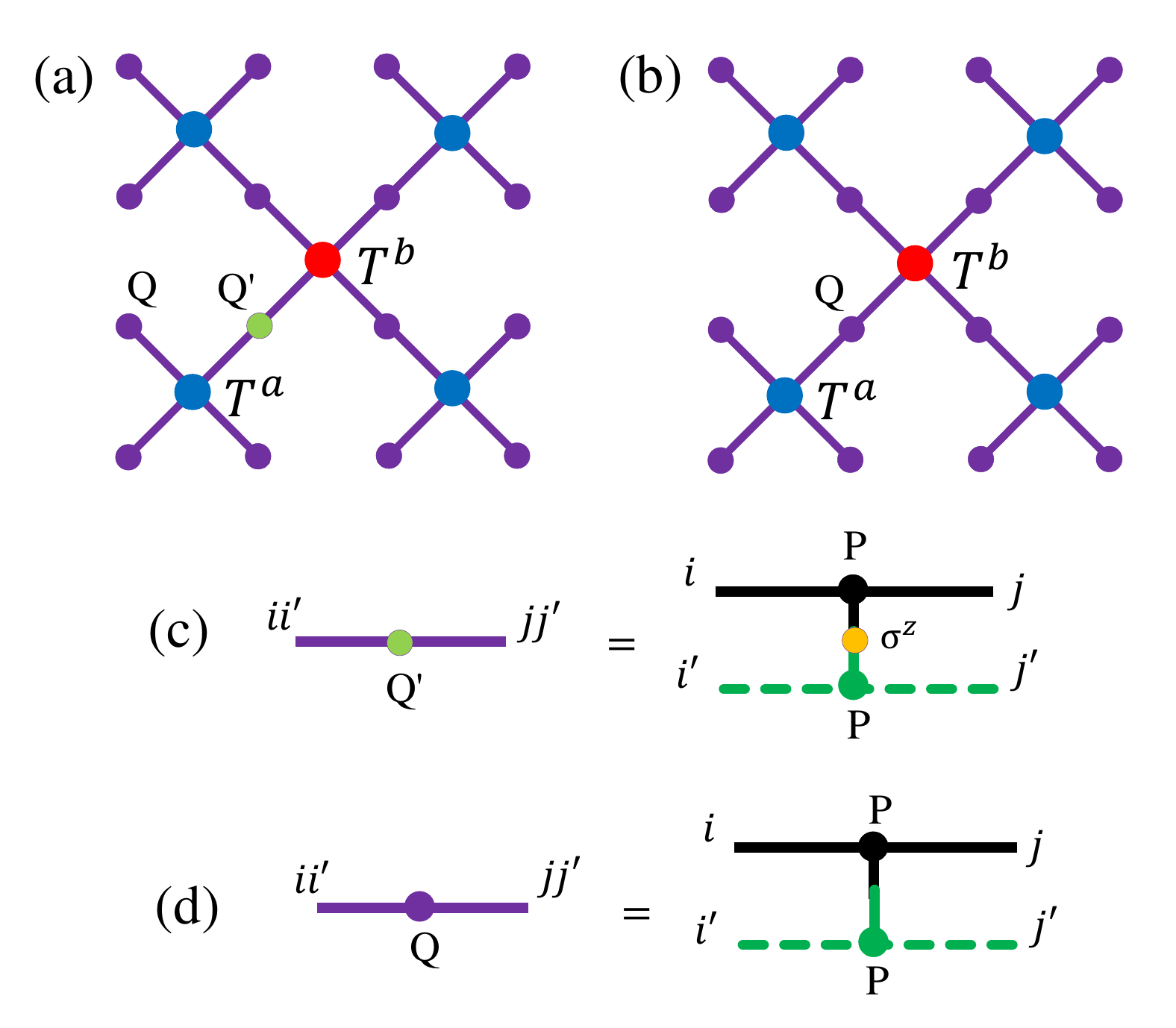}
	
	\caption{Sketch of the impurity method and the explanation of Eq.~(\ref{eq:app}). (a) The numerator $\langle \Psi'|s_i|\Psi\rangle$ represented as a 2D tensor network. (b) The denominator $\langle \Psi'|\Psi\rangle$ represented as a 2D tensor network. Note that almost all the tensors in (a) and (b) are the same, as defined in Fig.~\ref{fig:RTN}, and the only difference lies in two tensors defined on the $i$th site, i.e., $Q'$ (impurity tensor, denoted as bright green) as defined in (c) and $Q$ as defined in (d). In the definition of $Q'$, the operator $s_i$ is included, and here its matrix representation $\sigma^z$ is a diagonal matrix with nonzero elements $\pm 1$.}
	\label{fig:Impurity}
\end{figure}

The impurity method is frequently used to calculate the expectation values without resorting to the derivatives. It has been successfully applied early in Refs.~\cite{SU2D2008, SRG2009}, and was elaborated later in Refs.~\cite{HHZ-PRB2010, HHZ-PRB2016}. In essence, it aims to calculate the expectation values of some local operators, e.g., the bond energy or the magnetization, which can be expressed as a ratio of two quantities represented as tensor networks and differing from with each other by only very few tensors (dubbed impurity tensors). For simplicity, let us take the on-site magnetization operator $M_i$ as an example, and repeat Eq.~(\ref{SponM}) in the following
\begin{eqnarray}
	M_i = \frac{\langle \Psi'|s_i|\Psi\rangle}{\langle \Psi'|\Psi\rangle} \label{eq:app}
\end{eqnarray}
where $\langle\Psi'|$ and $|\Psi\rangle$ are the left and right dominant eigenvectors of $T_2T_1$ respectively. Suppose the two eigenvectors have been represented as PESS form, as shown in Fig.~\ref{fig:PESS}, then the numerator and the denominator can be represented as Figs.~\ref{fig:Impurity}(a) and \ref{fig:Impurity}(b) respectively. It shows that there is only one special impurity tensor $Q'$ in Fig.~\ref{fig:Impurity}(a), which constitutes the only difference from Fig.~\ref{fig:Impurity}(b). Thus $Q$ and $Q'$ share exactly the same effective environment $N^{(e)}$, and one can evaluate the ratio $M_i$ by determining $N^{(e)}$ through well-developed methods, e.g., the CTMRG method as described above, and obtain the following simple ratio form of $M_i$,
\begin{eqnarray}
	M_i = \frac{\mathrm{Tr}\left(N^{e}Q'\right)}{\mathrm{Tr}\left(N^{e}Q\right)}
\end{eqnarray} 
finally. For other local operators like bond energy, the process is very similar but involves two impurity tensors. More details and examples can be found in Ref.~\cite{HHZ-PRB2016}.

\subsection*{4. The result comparison between the CTMRG with and without nesting technique}
\begin{figure}[htbp]
	\includegraphics[width=0.48\textwidth]{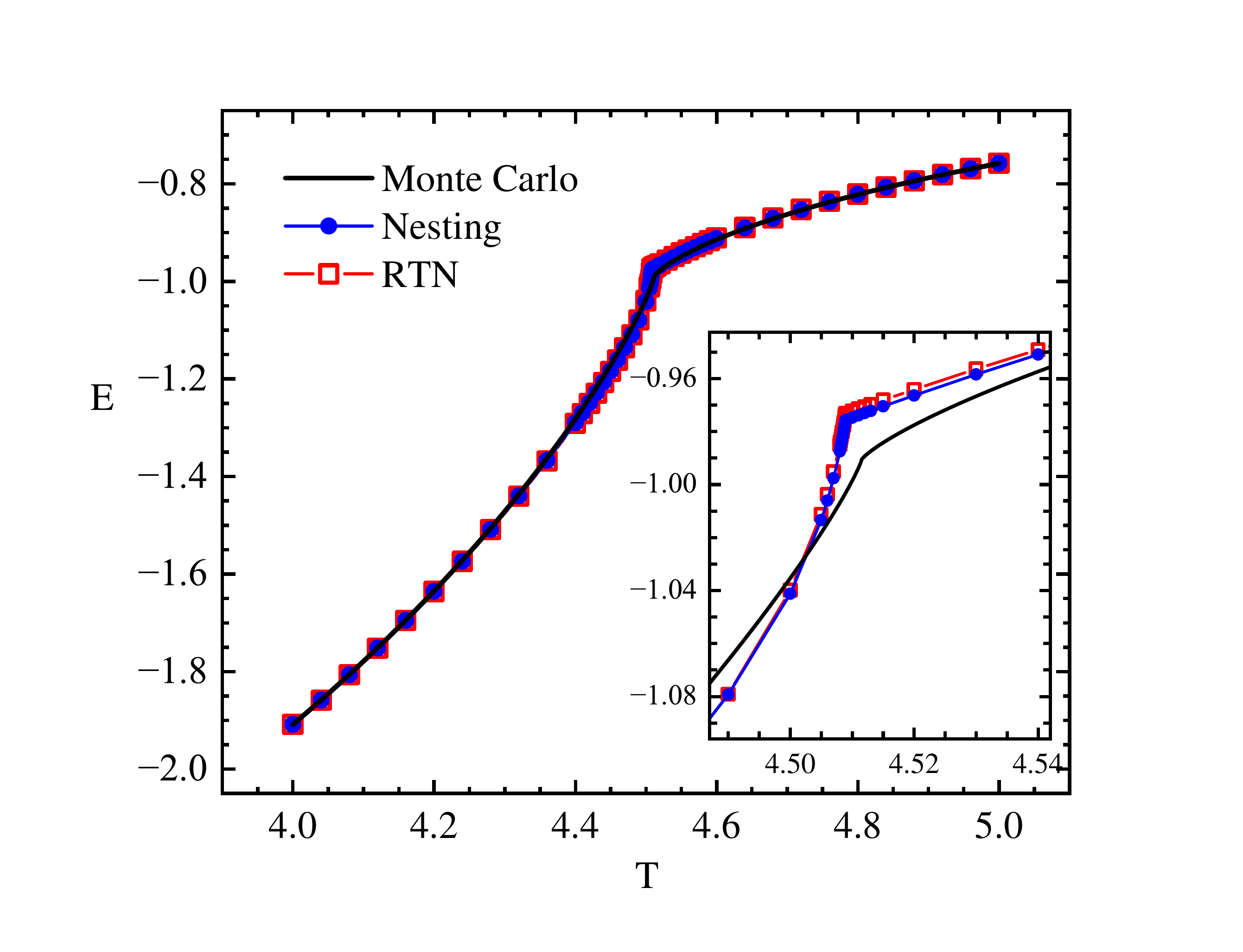}
	\caption{Comparison between the internal energy $E$ obtained from the conventional RTN method and the nesting technique, for $D=10$. Note that the PESS wave functions used in these two methods are the same, and in both methods, the network is contracted by the CTMRG algorithm.}
	\label{fig:CmpEnergy}
\end{figure}

\begin{figure}[htbp]
	\includegraphics[width=0.48\textwidth]{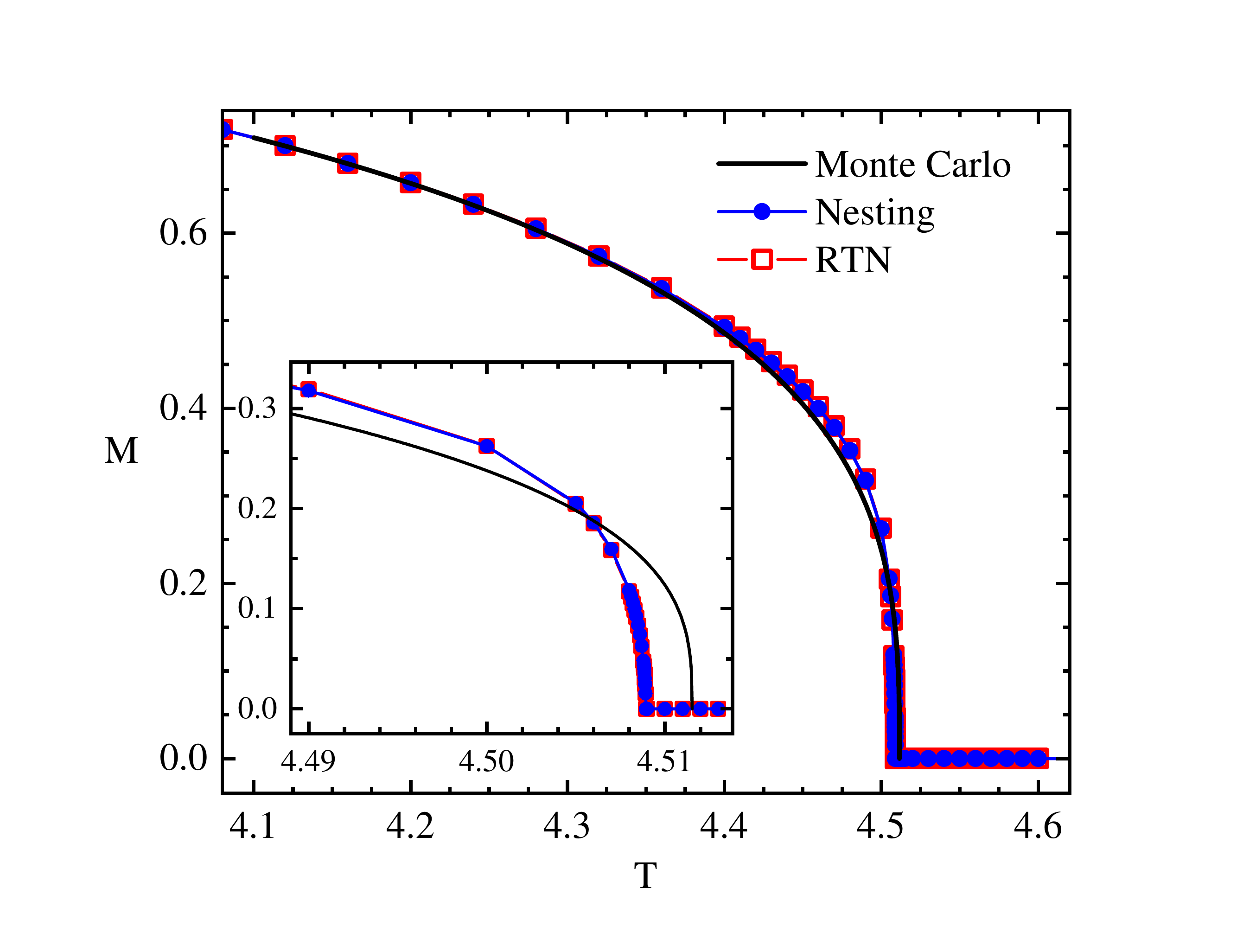}
	\caption{Comparison between the magnetization $M$ obtained from the conventional RTN method and the nesting technique, for $D=10$. Note that the PESS wave functions used in these two methods are the same, and in both methods, the network is contracted by the CTMRG algorithm.}
	\label{fig:CmpMag}
\end{figure}

In this paper, the data shown in the figures in the main text are all obtained by the CTMRG algorithms combined with the nesting technique. When $D$ is small and the conventional reduced tensor network method, i.e., RTN, can be efficiently performed, we can compare the 
results obtained by the two methods. The results are shown in Figs.~\ref{fig:CmpEnergy} and \ref{fig:CmpMag}. In both figures, complete convergence of the CTMRG calculations concerning the environment dimensions $\chi$ has been reached for both methods. It shows that both the internal energy $E$ and $M$ are very consistent for the two methods. Even though it seems that near critical region, the energy result obtained by the nesting technique is more closer to the Monte Carlo data, they are essentially the same. Actually the tiny difference comes not from the two methods themselves [i.e., whether use Fig.~\ref{fig:RTN} or Fig.~\ref{fig:NTN}(c) to represent the overlap of two PESSs, both of which are exact for two given representations], but from the fact that the methods employ different decompositions of the partition function [i.e., Fig.~\ref{fig:PF} and Eq.~(\ref{eq:PFCube}) in RTN, while Fig.~\ref{fig:TM4} and Eq.~(\ref{eq:Znest}) in the nesting technique], and that accordingly the two expectation value calculations involve different tensor networks. This tiny difference is expected to vanish gradually as $D$ becomes larger, and should vanish in quantum lattice models where exact Fig.~\ref{fig:TM4} and Eq.~(\ref{eq:Znest}) are unnecessary when employing the nesting technique.

\begin{thebibliography}{99}
	
\bibitem{PEPS2004}
F. Verstraete and J. I. Cirac, arXiv:cond-mat/0407066 (2004); J. I. Cirac, D. Perez-Garcia, N. Schuch, and F. Verstraete, Rev. Mod. Phys. \textbf{93}, 045003 (2021).

\bibitem{SimBook2018} 
S. Montangero, \textit{Introduction to Tensor Network Methods}, Springer (2018).

\bibitem{OrusReview} 
R. Orus, Ann. Phys. \textbf{349}, 117 (2014); R. Orus, Eur. Phys. J. B \textbf{87}, 280 (2014); R. Orus, Nat. Rev. Phys. \textbf{1}, 538 (2019).	
		
\bibitem{tJ2014} 
P. Corboz, T. M. Rice, and M. Troyer, Phys. Rev. Lett. \textbf{113}, 046402 (2014).

\bibitem{Simons-XXS}
Simons Collaboration on the Many-Electron Problem, Phys. Rev. X 5, 041041 (2015), and Phys. Rev. X \tbib{10}, 031016 (2020); 
B. X. Zheng, et al., Science \tbib{358}, 1155 (2017).

\bibitem{Corboz-PRL2014}
P. Corboz and F. Mila, Phys. Rev. Lett. \textbf{112}, 147203 (2014); P. Corboz and F. Mila, Phys. Rev. B \tbib{87}, 115144 (2013).

\bibitem{WL-PRB2016}
L. Wang, Z. C. Gu, F. Verstraete, and X. G. Wen, Phys. Rev. B \tbib{94}, 075143 (2016).

\bibitem{Liao-PRL2017}
H. J. Liao, Z. Y. Xie, J. Chen, Z. Y. Liu, H. D. Xie, R. Z. Huang, B. Normand, and T. Xiang, Phys. Rev. Lett. \textbf{118}, 137202 (2017)

\bibitem{LQ-PRB2022}
Q. Li, H. Li, J. Z. Zhao, H. G. Luo, and Z. Y. Xie, Phys. Rev. B \tbib{105}, 184418 (2022).

\bibitem{Ferrari-PRB2022} G. Ferrari, G. Magnifico, and S. Montangero, Phys. Rev. B\textbf{105}, 214201 (2022). 

\bibitem{HOTRG-PRB2012} 
Z. Y. Xie, J. Chen, M. P. Qin, J. W. Zhu, L. P. Yang, and T. Xiang, Phys. Rev. B \tbib{86}, 045139 (2012).


\bibitem{CW-PRB2014} 
C. Wang, S. M. Qin, and H. J. Zhou, Phys. Rev. B \textbf{90}, 174201 (2014).

\bibitem{ZP-PRL2021}
J. G. Liu, L. Wang, and P. Zhang, Phys. Rev. Lett. \tbib{126}, 090506 (2021).

\bibitem{Eisert-PRB2017}
C. Wille, O. Buerschaper, and J. Eisert, Phys. Rev. B \tbib{95}, 245127 (2017).

\bibitem{ZGM-PRL2020}
W. T. Xu, Q. Zhang, and G. M. Zhang, Phys. Rev. Lett. \tbib{124}, 130603 (2020).

\bibitem{Frank-arXiv2017}
D. J. Williamson, N. Bultinck, and F. Verstraete, arXiv:1711.07982.

\bibitem{RW-PBL2022}
R. Wang, Z. Y. Xie, B. G. Wang, and T. Sedrakyan, Phys. Rev. B \textbf{106}, L121117 (2022).

\bibitem{CMPS2010} 
F. Verstraete and J. I. Cirac, Phys. Rev. Lett. \textbf{104}, 190405 (2010).

\bibitem{LQCD2013} 
Y. Z. Liu, Y. Meurice, M. P. Qin, J. Unmuth-Yockey, T. Xiang, Z. Y. Xie, J. F. Yu, and H. Y. Zou, Phys. Rev. D \textbf{88}, 056005 (2013).

\bibitem{CTNS2019} 
A. Tilloy and J. I. Cirac, Phys. Rev. X \textbf{9}, 021040 (2019).


\bibitem{ML2018} 
Z. Y. Han, J. Wang, H. Fan, L. Wang, and P. Zhang, Phys. Rev. X \textbf{8}, 031012 (2018).

\bibitem{ML2020} 
Z. F. Gao, S. Cheng, R. Q. He, Z. Y. Xie, H. H. Zhao, Z. Y. Lu, and T. Xiang, Phys. Rev. Research \textbf{2}, 023300 (2020).

\bibitem{Yannick-RMP2022}
Y. Meurice, R. Sakai, and J. Unmuth-Yockey, Rev. Mod. Phys. \tbib{94}, 025005 (2022).
	
\bibitem{GV-TEBD}
G. Vidal, Phys. Rev. Lett. \tbib{91}, 147902 (2003), Phys. Rev. Lett. \tbib{93}, 040502 (2004).

\bibitem{SU1D2007} 
G. Vidal, Phys. Rev. Lett. \textbf{98}, 070201 (2007).

\bibitem{PESS2014} 
Z. Y. Xie, J. Chen, J. F. Yu, X. Kong, B. Normand, and T. Xiang, Phys. Rev. X \textbf{4}, 011025 (2014).

\bibitem{HHZ-PRB2010}
H. H. Zhao, Z. Y. Xie, Q. N. Chen, Z. C. Wei, J. W. Cai, and T. Xiang, Phys. Rev. B \textbf{81}, 174411 (2010).

\bibitem{Orus-SU2}
P. Schmoll, S. Singh, M. Rizzi, and R. Orus, Ann. Phys. \textbf{419}, 168232 (2020); M. Mambrini, R. Orus, and D. Poilblanc, Phys. Rev. B \textbf{94}, 205124 (2016).

\bibitem{RGMC-WHZ}
L. Wang, I. Pizorn, and F. Verstraete, Phys. Rev. B \textbf{83}, 134421 (2011); W.-Y. Liu, S.-J. Dong, Y.-J. Han, G.-C. Guo, and L. X. He, Phys. Rev. B \textbf{95}, 195154 (2017); H.-H. Zhao, K. Ido, S. Morita, and M. Imada, Phys. Rev. B \textbf{96}, 085103 (2017).

\bibitem{NTN-PRB2017}
Z. Y. Xie, H. J. Liao, R. Z. Huang, H. D. Xie, J. Chen, Z. Y. Liu, and T. Xiang, Phys. Rev. B \textbf{96}, 045128 (2017).

\bibitem{SachdevBook-QPT2011} S. Sachdev, \textit{Quantum Phase Transitions}, 2nd Edition, Cambridge University Press (2011).
\bibitem{Suzuki-PTP1976} M. Suzuki, Prog. Theo. Phys. \textbf{56}, 1454 (1976).
\bibitem{Nishino-JPSJ1995} T. Nishino, J. Phys. Soc. Jpn. \textbf{64}, 3598 (1995).
\bibitem{Xiang-JPCM1996} R. J. Bursill, T. Xiang, and G. A. Gehring. J. Phys.: Conden. Matt., \textbf{8}, L583 (1996).
\bibitem{Nig-ZPB1997} H. Niggemann, A. Klumper, and J. Zittartz, Z. Phys. B: Condens. Matter \textbf{104}, 103 (1997).

\bibitem{PYT-PA2017}
P. Y. Teng, Physica A \tbib{472}, 117 (2017).

\bibitem{Daisuke-arXiv2019}
D. Kadoh, and K. Nakayama, arXiv:1912.02414.

\bibitem{Daiki-PRB2020}
D. Adachi, T. Okubo, and S. Todo, Phys. Rev. B \tbib{102}, 054432 (2020).

\bibitem{Nishino-JPSJ1998}
T. Nishino and K. Okunishi, J. Phys. Soc. Jpn. \tbib{67}, 3066 (1998).

\bibitem{Nishino-KWA}
K. Okunishi and T. Nishino, Prog. Theor. Phys. \tbib{103}, 541 (2000); T. Nishino, K. Okunishi, Y. Hieida, N. Maeshima, and Y. Akutsu,
Nucl. Phys. B \tbib{575}, 504 (2000); A. Gendiar and T. Nishino, Phys. Rev. E \tbib{65}, 046702 (2002).

\bibitem{Nishino-TPVA}
T. Nishino, Y. Hieida, K. Okunishi, N. Maeshima, Y. Akutsu, and A. Gendiar, Prog. Theor. Phys. \tbib{105}, 409 (2001); A. Gendiar, N. Maeshima, and T. Nishino, Prog. Theor. Phys. \tbib{110}, 691 (2003).

\bibitem{SGC-PLA2006}
S. G. Chung, Phys. Lett. A \tbib{359}, 707 (2006).

\bibitem{Orus-CTTRG2012}
R. Orus, Phys. Rev. B \tbib{85}, 205117 (2012).

\bibitem{FV-PRE2018}
L. Vanderstraeten, B. Vanhecke, and F. Verstraete, Phys. Rev. E \tbib{98}, 042145 (2018).

\bibitem{LV-PRL2022}
B. Vanhecke, J. Hasik, F. Verstraete, and L. Vanderstraeten, Phys. Rev. Lett. \textbf{129}, 200601 (2022).
\bibitem{Nishino-JPSJ2022} K. Okunishi, T. Noshinio, and H. Ueda, J. Phys. Soc. Jpn. \textbf{91}, 062001 (2022).
\bibitem{SU2D2008} 
H. C. Jiang, Z. Y. Weng, and T. Xiang, Phys. Rev. Lett. \textbf{101}, 090603 (2008).

\bibitem{GV-NUTEBD}
Note that, essentially the similar idea has been applied to two-dimensional classical models under the help of the canonical form of matrix product state. For example, see R. Orus, and G. Vidal, Phys. Rev. B \textbf{78}, 155117 (2008).

\bibitem{CTMRG1996} 
T. Nishino and K. Okunishi, J. Phys. Soc. Jpn. \textbf{65}, 891 (1996).

\bibitem{CTMRG2009} 
R. Or\'us and G. Vidal, Phys. Rev. B \textbf{80}, 094403 (2009).

\bibitem{Schuch-PRB2012}
N. Schuch, D. Poilblanc, J. I. Cirac, and D. P. Garcia, Phys. Rev. B \tbib{86}, 115108 (2012).


\bibitem{HHZ-PRB2016}
H. H. Zhao, Z. Y. Xie, T. Xiang, and M. Imada, Phys. Rev. B \textbf{93}, 125115 (2016).

\bibitem{SW-CPL2014}
S. Wang, Z. Y. Xie, J. Chen, B. Normand, and T. Xiang, Chin. Phys. Lett. \textbf{31}, 070503 (2014).

\bibitem{Blote-JPA1996}
A. L. Talapov and H. W. J. Blote, J. Phys. A: Math. Gen. \textbf{29}, 5727 (1996).

\bibitem{Blote-PRE}
Y. J. Deng and H. W. J. Blote, Phys. Rev. E \textbf{68}, 036125 (2003); X. M. Feng and H. W. J. Blote, Phys. Rev. E \textbf{81}, 031103 (2010).

\bibitem{Hasen-PRB2010}
M. Hasenbusch, Phys. Rev. B \textbf{82}, 174433 (2010).

\bibitem{Landau-PRE2018}
A. M. Ferrenberg, J. Xu, and D. P. Landau, Phys. Rev. E \textbf{97}, 043301 (2018).

\bibitem{SRG2009}
Z. Y. Xie, H. C. Jiang, Q. N. Chen, Z. Y. Weng, and T. Xiang, Phys. Rev. Lett. \textbf{103}, 160601 (2009).

\bibitem{WL-PRB2012}
W. Li, J. von Delft, and T. Xiang, Phys. Rev. B, \textbf{86}, 195137 (2012).

\bibitem{HJL-PRB2016}
H. J. Liao, Z. Y. Xie, J. Chen, X. J. Han, H. D. Xie, B. Normand, and T. Xiang, Phys. Rev. B, \textbf{93}, 075154 (2016).

\bibitem{TPVA-2} A. Gendiar, and T. Nishino, Phys. Rev. B \textbf{71}, 024404 (2005).

\bibitem{Hasen-IJMPC2001} M. Hasenbusch, Int. J. Mod. Phys. C \textbf{12}, 911 (2001).
\bibitem{Vichi-JHEP2016} F. Kos, D. Poland, D. S. Duffin, and A. Vichi, J. High Energ. Phys. \textbf{08}, 036 (2016) 
\bibitem{Sandvik-PRL2020}
H. Shao, W. Guo, and A. W. Sandvik, Phys. Rev. Lett. \textbf{124}, 080602 (2020).

\bibitem{CU-PRB2016}
L. Wang, and F. Verstraete, arXiv:1110.4362.

\bibitem{FU-PRL2008}
J. Jordan, R. Orus, G. Vidal, F. Verstraete, and J. I. Cirac, Phys. Rev. Lett. \textbf{101}, 250602 (2008);

\bibitem{FU-PRB2014}
M. Lubasch, J. I. Cirac, and M. C. Banuls, Phys. Rev. B \textbf{90}, 064425 (2014);
H. N. Phien, J. A. Bengua, H. D. Tuan, P. Corboz, and R. Orus, Phys. Rev. B \textbf{92}, 035142 (2015).

\bibitem{LXC-arXiv2022}
L. X. Cen, arXiv:2208.03436.

\bibitem{AA-arXiv2022}
A. Acuaviva, V. Makam, H. Nieuwboer, D. P. Garcia, F. Sittner, M. Walter, and F. Witteveen, arXiv:2209.14358.

\bibitem{SSL-From1981}
B. S. Shastry and B. Sutherland, Physica \textbf{108B}, 1069 (1981); J. Yang, A. W. Sandvik, and L. Wang, Phys. Rev. B \textbf{105}, L060409 (2022).

\bibitem{HYZ-arXiv2020}
H. Y. Zou, F. Yang, and W. Ku, arXiv:2011.06520.











\bibitem{YJK-PRB2018} C. Y. Lee, B. Normand, and Y. J. Kao, Phys. Rev. B \textbf{98}, 224414 (2018).
\bibitem{Sen-PRB2022} S. Niu, J. Hasik, J. Y. Chen, and D. Poilblanc, Phys. Rev. B \textbf{106}, 245119 (2022).
\bibitem{LY-PRB2023} Y. Liu, Z. Y. Xie, H. G. Luo, and J. Z. Zhao, Phys. Rev. B \textbf{107}, L041106 (2023). 
\bibitem{XFL-CPL2022} X. F. Liu, Y. F. Fu, W. Q. Yu, J. F. Yu, and Z. Y. Xie,  Chin. Phys. Lett. \textbf{39}, 067502 (2022).
\bibitem{Lukin-PRB2023} I. V. Lukin, and A. G. Sotnikov, Phys. Rev. B \textbf{107}, 054424 (2023).
\end{thebibliography}
\end{document}